\documentclass{aa}  

\usepackage{graphicx}
\usepackage{txfonts}
\usepackage{natbib}
\bibpunct{(}{)}{;}{a}{}{,} 

\usepackage{color}
\definecolor{darkgreen}{RGB}{0,142,128}

\usepackage{hyperref}
\hypersetup{colorlinks,citecolor=blue,linkcolor=blue}

\newcommand{\modif}[1]{{{{#1}}}}

\usepackage{multirow}

\defcitealias{cohen_heating_2024}{C24}

\usepackage{placeins}

\begin{document}

   \title{Ohmic heating in the upper atmosphere of hot exoplanets}
   \subtitle{The influence of a time-varying magnetic field}

   \author{A. Strugarek
          \inst{1}
          \and
          A. Garc\'ia Mu{\~n}oz\inst{1}
          \and 
          A. S. Brun\inst{1}
          \and 
          A. Paul\inst{1}
          }

   \institute{D\'epartement d'Astrophysique/AIM
CEA/IRFU, CNRS/INSU, Univ. Paris-Saclay \& Univ. de Paris
91191 Gif-sur-Yvette, France\\
              \email{antoine.strugarek@cea.fr}
              }

   \date{}

 
  \abstract
  {Exoplanets on close-in orbit are subject to intense X-ray and ultraviolet (XUV) irradiation from their star. Their atmosphere therefore heats up, sometimes to the point where it thermally escape from the gravitational potential of the planet. Nonetheless, XUV is not the only source of heating in such atmospheres. Indeed, close-in exoplanets are embedded in a medium (the stellar wind) with strong magnetic fields that can significantly vary along the orbit. The variations of this magnetic field can induce currents in the upper atmosphere, which dissipate and locally heat it up through Ohmic heating.}
   {The aim of this work is to quantify Ohmic heating in the upper atmosphere of hot exoplanets due to an external time-varying magnetic field, and to compare it to the XUV heating.}
   {Ohmic heating depends strongly on the conductivity properties of the upper atmosphere. A 1D formalism is developed to assess the level and the localization of Ohmic heating depending on the conductivity profile. 
   The formalism is applied to the specific cases of Trappist-1 b and $\pi$ Men c.}
   {Ohmic heating can reach values up to 10$^{-3}$ erg s$^{-1}$ cm$^{-3}$ in the upper atmospheres of hot exoplanets. It is expected to be stronger the closer the planet is and the lower the central star mass is, as these conditions maximize the strength of the ambient magnetic field around the planet. The location of maximal heating depends on the conductivity profile (but does not necessarily occurs at the peak of conductivity), and in particular on the existence and strength of a steady planetary field. Such extra heating can play a role in the thermal budget of the escaping atmosphere 
   \modif{when the planetary atmospheric magnetic fields is comprised between} 0.01 G and 1G.}
   {We confirm that Ohmic heating can play an important role in setting the thermal budget of the upper atmosphere of hot exoplanets, and can even surpass the XUV heating in the most favorable cases. When it is strong, a corollary is that the upper atmosphere screens efficiently time-varying external magnetic fields, preventing them to penetrate deeper in the atmosphere or inside the planet itself. We find that both Trappist-1b and $\pi$ Men c are likely subject to intense Ohmic heating. 
   }

   \keywords{Planet-star interactions -- Planets and satellites: magnetic fields -- Planets and satellites: atmospheres }

   \maketitle
%

\section{Introduction}

Ohmic heating is a physical process where the dissipation of \modif{electric currents} leads to a local heating of the medium. It is linked to the dissipative properties of the medium, namely its conductivity. The latter is \modif{determined} by the atomic and molecular composition of the medium, its temperature, its ionization level and the existence \modif{(or non-existence) of steady large scale magnetic fields}. In the context of exoplanets, Ohmic heating has been invoked as a possible source of heating for the deep interior of planets (of any size) and for their deep atmosphere. Here, we consider its possible role in the upper atmosphere of exoplanets, {i.e.} in their ionosphere and above. In what follows, we will use "upper atmosphere" to mean the layers above the 1 $\mu$bar-level, which broadly coincides with the ionosphere.

Ohmic heating requires the existence of \modif{an electric current} in the first place. This \modif{current} can be produced \modif{by various means}, for instance \modif{due to electric potential differences as in the Earth magnetosphere, or} by the shearing of a pre-existing magnetic field due to ionized zonal winds. This \modif{latter} specific case has been proposed to be acting in the deep atmosphere of hot Jupiters \citep[{e.g.}][and references therein]{batygin_inflating_2010,huang_ohmic_2012,rogers_magnetic_2014}, where zonal winds can shear a dipolar magnetic field sustained by dynamo action in the planet interior. In that context, the conductivity level and its anisotropies (e.g. day-night) determine largely the Ohmic heating rate, and whether it contributes to the thermal budget of the atmosphere or not \citep[{e.g.}][]{dietrich_magnetic_2022}.

Nevertheless, other processes can lead to the existence of \modif{electric currents}. For instance, the motion of a planet along its orbit can lead to spatial and temporal variations of the stellar wind magnetic field as seen from the planet. This naturally occurs due to the complexity of stellar magnetic field in their photosphere and above, that generally shapes the magnetic environment of stars and the stellar wind magnetic field intercepted by planets along their orbit. In addition, planets with eccentric orbits will generally probe various magnetic field strengths along their orbit as well. \modif{In these cases, the variations of the magnetic field around the planet can induce electric currents in the planet itself}. \citet{chyba_internal-current_2021} studied the particular case of induction and Ohmic heating associated with a spatially-variable magnetic field permeating a planetary interior with a variable conductivity profile, which could be at play within the \modif{jovian satellite} Io. The case of temporal variation of an external field has been explored further in the context of solar system planets and satellites, as well as exoplanets. Induction due an external magnetic field could be for instance at the origin of zonal jets in the interior of Europa \citep{Gissinger2019}, but is thought to be generally ineffective in heating the interior of solar system satellites \citep{chyba_magnetic_2021}. In the context of exoplanets, close-in planets interact with a much stronger magnetic field than solar system planets \citep{strugarek_interactions_2024}. If this magnetic field exhibits temporal variations, \citet{kislyakova_magma_2017,kislyakova_electromagnetic_2020} proposed it could lead to significant additional heating in their interior, possibly triggering volcanic activity. This process will be maximized naturally around planet-hosting stars producing strong magnetic fields \citep{kislyakova_effective_2018}, which are often young, fast rotating, low mass stars \citep[{e.g.}][]{morin_large-scale_2010,reiners_magnetism_2022}. 

The same process can also, in principle, occur in the upper atmosphere of exoplanets, as we will demonstrate in this work. In upper planetary atmospheres, the conductivity is generally anisotropic due to the presence of a magnetic field. Indeed, at these altitudes the gyrofrequency of electrons tends to be larger than their collision frequency with the surrounding ion and/or neutral species, changing drastically the effective conductivity parallel and perpendicular to the local magnetic field \citep[{e.g.} see section 5.11 in][]{schunk_ionospheres_2009}. In addition, hot exoplanets generally harbor large numbers of electrons in their upper atmosphere due to the ionization by the strong XUV flux they receive. This leads to large conductivities at these altitudes, \modif{enabling the dissipation of the currents induced by the external time-varying field.} 
In this context, a first study has been conducted by \citet{cohen_heating_2024}, predicting very high heating rates for conductivities prescribed {a priori} \modif{(a detailed comparison to this work is shown in Appendix \ref{sec:C24_comparison}}). In this work, we go beyond this approach and used ab-initio models of the upper atmosphere of exoplanets \citep{garcia_munoz_heating_2023} to assess the conductivities \citep[{e.g.}][]{Johnstone2018} and associated Ohmic heating. 

Several questions arise: which planets have an atmosphere that is unaffected by such a time-varying magnetic field? For the others, if a planet screens it, can the associated Ohmic heating change the energy balance of its upper atmosphere? To answer these questions, we use a generic formalism for the Ohmic heating associated with an external time-varying magnetic field to assess the penetration of such fields through the upper atmosphere of hot exoplanets. We develop the formalism based on the classical theory of \citet{Parkinson} in Sect.  \ref{sec:formalism}. Then, in Sect. \ref{sec:IllustrativeExamples} we propose examples to illustrate how changes in the conductivity and in the oscillation frequency of the external magnetic field affect the penetration and heat deposition. We then apply the formalism to realistic upper atmosphere models of Trappist-1 b and $\pi$ Men c in Sect. \ref{sec:T1b_PMc}. Then, we generalize our approach to provide constraints on the star-planet systems susceptible to lead to strong Ohmic heating in their upper atmosphere in Sect. \ref{sec:exoplanetPop}. We discuss the implications of our results in Sect. \ref{sec:discussion} and conclude in Sect. \ref{sec:conclusion}.

\section{Penetration of a time-varying magnetic field in a planetary atmosphere and Ohmic heating}
\label{sec:formalism}


\subsection{Physical-mathematical formulation of the problem}
\label{sec:problem_formulation}

Let us consider a plasma composed of neutrals (denoted $n$), ions (denoted $i$) and electrons (denoted $e$). We will assume that the ions are singly ionized, which is typical for the upper atmosphere of metal-poor exoplanets, such that the total ions and electron number densities are equal ($n_i=n_e$). 

The general Ohm's law for charged species $\alpha$ (electrons or ions) can be written as \citep{norman_anomalous_1985}:
\begin{equation}
    {\bf J}_\alpha = \underline{\sigma}_\alpha \cdot {\bf E}\, ,
\end{equation}
where ${\bf J}_\alpha$ is the current density carried by charged species $\alpha$, $\underline{\sigma}_\alpha$ is the conductivity tensor associated with species $\alpha$ and ${\bf E}$ is the electric field. The total current density can then be expressed as 
\begin{equation}
    {\bf J} = \sum_\alpha {\bf J}_\alpha \, .
\end{equation}
In addition, Ampere's law can be written as (in cgs)
\begin{equation}
    \boldsymbol{\nabla}\times{\bf B} = \frac{4\pi}{c} {\bf J}\, .
\end{equation}
Because $\boldsymbol{\nabla}\cdot{\bf B}=0$, we can introduce the magnetic vector potential ${\bf A}$ such that $\boldsymbol{\nabla}\times{\bf A} = {\bf B}$. By definition, ${\bf A}$ is defined up to a gradient $\nabla \phi$. Putting ourselves in the particular gauge where $\boldsymbol{\nabla}\cdot{\bf A}=0$ ({i.e.} $\nabla^2 \phi=0$), we obtain that
\begin{equation}
    -\nabla^2 {\bf A} = \frac{4\pi}{c} \sum_\alpha \underline{\sigma}_\alpha \cdot {\bf E}\, .
\end{equation}
In addition, Faraday's law stipulates that $c{\bf E} = -\partial_t {\bf A}$, therefore
\begin{equation}
    \label{eq:MasterEq}
    \nabla^2 {\bf A} - \frac{4\pi}{c^2}\sum_\alpha \underline{\sigma}_\alpha \cdot  \partial_t {\bf A} = 0 \, .
\end{equation}

This equation is the fundamental equation that was solved e.g. by \citet{Parkinson} for the induction and heating within a planet, \modif{reviewed by \citet{Saur2010} for general planetary bodies,} and later applied to different types of exoplanetary interior by \citet{kislyakova_magma_2017}. The formalism is general for any conducting medium, and an analytical solution can be found when the conductivity is piece-wise constant in space \citep{Parkinson}. Here, we will opt to solve it numerically in 1D to take into account arbitrary profiles of conductivity. We have validated the implementation of the solver against the analytical solution in the case of a constant conductivity, which is derived in Appendix \ref{sec:constantPedersenCondCase}.

We will furthermore assume as a first approximation that the time-varying component can be approximated by a sinusoidal variation of frequency $\Omega$ such that 
\begin{equation}
    {\bf A} = {\bf A}_{\rm sw}({\bf x},t) + {\bf A}_{P}(x)\, , 
\end{equation}
where 
\begin{equation}
    {\bf A}_{\rm sw}({\bf x},t) = {\bf A}_0 ({\bf x}) e^{-i\Omega t}\, ,
\end{equation}
and where ${\bf A}_{P}({\bf x})$ is the steady magnetic field. We note that in principle, ${\bf A}_{P}({\bf x})$ is the sum of a steady component from the stellar wind and of a steady planetary magnetic field. Because we assume that we are within the magnetosphere of the planet here, it is natural to assume that the steady component of the stellar wind is small compared to the steady planetary magnetic field and therefore to neglect it.

It must be noted that the conductivity tensor $\underline{\sigma}_\alpha$ in Eq. (\ref{eq:MasterEq}) also formally depends on the magnetic field, and therefore depends on time. Integrating \ref{eq:MasterEq} over an oscillation period ($\int_{0}^{P} . e^{i\Omega t} {\rm d}t$, with $P=2\pi/\Omega$) makes the parts of the conductivity tensor that depend on time vanish, which results into 
\begin{equation}
    \nabla^2 {\bf A}_0 + \frac{4\pi i \Omega}{c^2} \sum_\alpha \underline{\sigma}_\alpha({\bf x}) \cdot  {\bf A}_0 = 0 \, .
\end{equation}

\begin{figure}[!htbp]
    \centering
    \includegraphics[width=\linewidth]{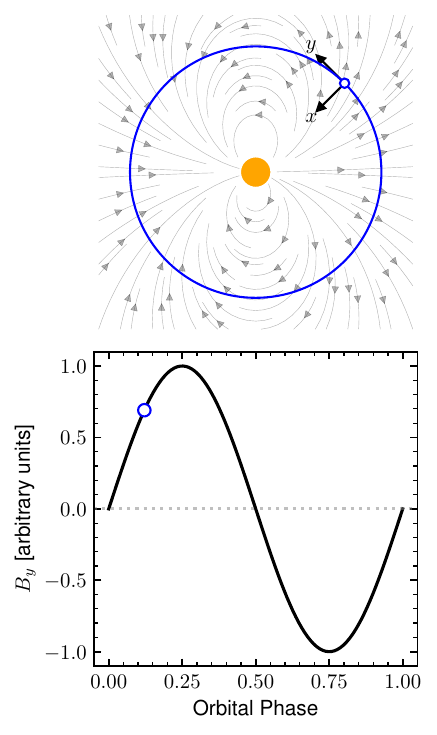}
    \caption{Top panel: schematic of the geometry considered in this work. The central star (orange circle) creates a dipole field (gray arrows and lines) in which the planet (blue open circle symbol) orbits (large blue circle). The $x$ axis corresponds to the star-planet direction at all times, and the $y$ axis the direction of the orbit. The $z$ axis (not shown here) is perpendicular to the orbital plane. Bottom panel: the time-varying magnetic field $B_y$ (in arbitrary unit) seen by the planet along its orbit. The open blue circle corresponds to the orbital phase illustrated in the top panel. This schematic can either represent the case of an inclined orbit with a stellar dipole aligned with the stellar rotation axis, or the case of a star sustaining a dipole inclined by 90$^\circ$ with respect to its rotation axis.}
    \label{fig:schematic}
\end{figure}

Solving this equation in a domain with a space-varying conductivity allows assessing both (i) how much Ohmic heating can be deposited due to the existence on an external time-varying magnetic field, (ii) where the Ohmic heating can be deposited, and (iii) how much of the time-varying magnetic field can permeate below the conducting region. In this work, we aim to characterize these three points for the atmosphere of hot exoplanets. 

To do so, we will consider a simplified geometry which is illustrated in Fig. \ref{fig:schematic}. This geometry can be thought to be representative of various situations. It can represent a case where the star exhibits a dipolar field aligned with the rotation axis (roughly like the Sun during activity minima), and where the planet is on a 'polar' orbit. Close-in planets such as GJ 436b harbor such orbits \citep{bourrier_polar_2022}, and could be statistically abundant \citep{albrecht_preponderance_2021}. Our simplified geometry could equally represent classical orbits perpendicular to the stellar rotation axis, with a star harboring an inclined dipole (like, for instance, many M-dwarfs \citealt{morin_large-scale_2010}). We consider that the time-varying magnetic field is aligned with the local Cartesian axis ${\bf e}_y$. The Cartesian axis ${\bf e}_x$ is the star-planet axis, we will consider that variables can vary in time, and but can vary in space only along the $x$ axis as we are considering a 1D model.

In addition, we will assume that the steady magnetic field (here assumed to be mostly of planetary origin) is aligned with the ${\bf e}_z$ axis. If the magnetic field is strong enough, the conductivity in the medium becomes anisotropic \citep[{e.g.}][]{schunk_ionospheres_2009}. In that case, given the simplified geometry we chose, the conductivity tensor reduces to
\begin{equation}
\underline{\sigma}_\alpha = \left( 
    \begin{array}{ccc}
     \sigma_{P,\alpha} & \sigma_{H,\alpha}  & 0 \\
    -\sigma_{H,\alpha} & \sigma_{P,\alpha} & 0 \\
     0 & 0 & \sigma_{0,\alpha}
    \end{array}
\right) ,
\end{equation}
where we have introduced the parallel, field-aligned conductivity $\sigma_0$, the Hall conductivity $\sigma_H$ and the Pedersen conductivity $\sigma_P$. The total conductivities are given by the sum of the individual conductivities associated with species $\alpha$, such that 
\begin{eqnarray*}
    \sigma_0 &=& \sum_\alpha \sigma_{0,\alpha} \, , \\ 
    \sigma_H &=& \sum_\alpha \sigma_{H,\alpha} \, ,\\
    \sigma_P &=& \sum_\alpha \sigma_{P,\alpha} \, .
\end{eqnarray*}
We note that with the simplified geometry we have assumed here, the vector potential can be written as ${\bf A}_0(x) = A_0(x){\bf e}_y$. As a result, Eq. (\ref{eq:MasterEq}) becomes
\begin{equation}
\label{eq:FinalMasterEq_tmp}
    \partial_{xx} A_0 + \frac{4\pi i \Omega \sigma_{P}}{c^2} A_0 = 0\, .
\end{equation}
This equation leads to the introduction of the well-known skin-depth \citep[{e.g.}][]{Parkinson}
\begin{equation}
\label{eq:SkinDepth}
    \delta_{P}  = \frac{c}{\sqrt{2\pi \Omega \sigma_{P}}} \, 
\end{equation}
to simplify equation (\ref{eq:FinalMasterEq_tmp}) and obtain 
\begin{equation}
\label{eq:FinalMasterEq}
    \partial_{xx} A_0 +  \frac{2i}{\delta_P^2} A_0 = 0\, .
\end{equation}

In what follows, we will solve Eq. (\ref{eq:FinalMasterEq}) for space-varying conductivity profiles and for an external time-varying magnetic field. To do so, we will consider an atmosphere of a certain depth subject to the following boundary conditions. At the top of the atmosphere, we will consider that the time-varying field is forced at the external boundary to be $B_{\rm sw}$, which translates into a Neumann boundary condition on $A_0$ such that $\partial_x A_0 = B_{\rm sw}$. At the bottom boundary condition, we will consider that the current density is zero which means that $\partial_{xx}A_0=0$. These boundary conditions leads to families of solutions for $A_0$ that can be obtained up to an additive constant. This is not problematic, since we are interested in characterizing here only the penetration of the ambient time-varying magnetic field and the heat deposition associated with the triggered currents. Both aspects are independent of the chosen additive constant when solving Eq. (\ref{eq:FinalMasterEq}) to obtain $A_0$. 

After having solved Eq. (\ref{eq:FinalMasterEq}), the volumetric heat deposition in the atmosphere can be assessed through $Q = {\bf J}\cdot{\bf E}$ which translates into
\begin{eqnarray}
\label{eq:heating}
    Q &=&  {\bf J} \cdot (\underline{\sigma}^{-1} {\bf J})  = \frac{\sigma_P}{\sigma_H^2+\sigma_P^2} J_y^2  = \frac{\sigma_P}{\sigma_H^2+\sigma_P^2} \left(\frac{c}{4\pi} \Re (\partial_{xx} A_0)\right)^2 \nonumber \\ 
    &=& \frac{\sigma_P}{\sigma_H^2+\sigma_P^2} \left(\frac{c}{2\pi\delta_P^2} \Im (A_0)\right)^2 \, ,
\end{eqnarray} 
where we have made use of Eq. (\ref{eq:FinalMasterEq}) to simplify the final heating rate formula, and $\Re$ and $\Im$ respectively stand for the real and imaginary parts. Finally, we can also show (see Appendix \ref{sec:constantPedersenCondCase}) that in the limiting case where $\sigma_P$ is constant and $\delta_P$ is smaller than the vertical extent of the atmosphere considered, the heating rate becomes independent of $\sigma_p$ and can be considered as the maximal heating rate that can be achieved:
\begin{equation}
    \label{eq:maxQ_simplified}
    Q_{\rm max} = \frac{\Omega B_{\rm sw}^2}{8\pi}\, .
\end{equation}

\subsection{Assessment of conductivities}
\label{sec:ConductivityFormulae}

Following the approximation proposed by \citet{chapman_electrical_1956}, we can estimate the conductivities when considering singly-ionized particles through \citep{norman_anomalous_1985,Johnstone2018}
\begin{eqnarray}
\label{eq:sig0}
    \sigma_{0,\alpha} &=& \frac{e^2 n_\alpha }{m_\alpha \nu_\alpha} \, , \\ 
\label{eq:sigH}
    \sigma_{H,\alpha} &=&  \frac{\nu_\alpha\omega_\alpha}{\nu_\alpha^2 + \omega_\alpha^2}\sigma_{0,\alpha} \, ,\\
\label{eq:sigP}
    \sigma_{P,\alpha} &=& \frac{\nu_\alpha^2}{\nu_\alpha^2 + \omega_\alpha^2}\sigma_{0,\alpha} \, ,
\end{eqnarray}
where $n_\alpha$ is the number density of species alpha, $m_\alpha$ its mass, $\nu_\alpha$ its collision frequency with other species, and $\omega_\alpha= -q_\alpha B_{P} / (c m_\alpha)$ its algebraic gyrofreqency. We note several interesting limits to these formulae. First, if $\omega_\alpha$ is set to zero we obtain that $\sigma_{H,\alpha}=0$ and $\sigma_{P,\alpha}=\sigma_{0,\alpha}$, which makes the conductivity tensor symmetric. Another interesting limit exists in planetary atmospheres. In general, even for small magnetic fields of the order of a few nT (a few $10^{-5}$ G), it is found that $\nu_e \ll \omega_e$ in planetary atmospheres. This means that in the limit of small magnetic field, the Hall conductivity of electrons does not vanish in planetary atmospheres and shall always be considered in induction-heating applications.

This formulation of the conductivity tensor is generic. Further approximations have been proposed in the literature in the context of the Earth ionosphere, assuming some dominant species and collision frequencies \citep[{e.g.}][]{maeda_conductivity_1977}. We will not use these approximations here, to retain a generic formulation of the conductivity associated with each individual species. We nevertheless compare our generic formulation to the reduced formulation of \citet{maeda_conductivity_1977} in the case of the Earth in Appendix \ref{sec:EarthAppendix} and obtained a satisfying agreement,

We also note that the conductivity of a medium can take different expressions depending on whether it is subject to a direct or alternating current. Here, the current develops on the timescale of the time-varying magnetic field. In this study, this will typically be of the order of magnitude of the orbit of the planet, {i.e.} from tenths of a day to a few days. This timescale is much longer than all relevant collision timescales within the atmosphere (see Appendix \ref{sec:CollisionFreqsT1bPMc}). As a result, the atmosphere effectively sees a direct current on collision timescales, and the modeling considered here applies to the case of direct currents only. 

\subsection{Assessment of collision frequencies}
\label{sec:CollFreqs}

The collision frequencies $\nu_\alpha$ can be estimated based on the relative abundances of electrons/ions and neutrals in the atmosphere. We will assume in this study that all species have the same temperature, for the sake of simplicity. We make use of the parametrized collision frequencies summarized in \citet{schunk_ionospheres_1980}. The collisions of species $\alpha$ with species $\beta$ are parametrized through
\begin{equation}
    \nu_{\alpha\beta} = C_{\alpha\beta}(T) n_\beta\, ,
\end{equation}
where the collision coefficient $C_{\alpha\beta}$  depends on the local temperature and $n_\beta$ is the number density of species $\beta$. In \citet{schunk_ionospheres_1980}, the electrons-neutral collision coefficients are tabulated in their Table 3. The ion-neutral collision coefficients are tabulated in their Table 6 for non-resonant interactions and Table 5 for resonant interactions. The electron-ion collision frequencies are generically parametrized through their equation (15):
\begin{equation}
    \nu_{ei} = \frac{54.5}{T^{3/2}} n_i\,\,\,\, {\rm [s]} \, ,
\end{equation}
where we have considered only singly-ionized species, and where $T$ in expressed in [K] and $n_i$ in [g cm$^{-3}$]. Note that these collisions frequencies are such that 
\begin{equation*}
    n_\alpha m_\alpha \nu_{\alpha\beta} = n_\beta m_\beta \nu_{\beta\alpha}\, .
\end{equation*}
When $n_e=n_i$, this means that $\nu_{ie}=(m_e/m_i)\nu_{ei}$, which can be understood as an ion collides less often with electrons that an electron collides with ions.

Finally, to compute the conductivities introduced in Eqs. (\ref{eq:sig0}, \ref{eq:sigH}, \ref{eq:sigP}) we define the ion species and electron collision frequencies $\nu_\alpha$ ($\alpha \in (i,e)$) as
\begin{eqnarray*}
     \nu_i &=& \nu_{ie} + \sum_{\alpha \in (n)} \nu_{i\alpha} \, ,\\
     \nu_e &=& \sum_{\alpha\in (i,n)} \nu_{e\alpha} \, .
\end{eqnarray*}

\begin{figure*}
    \centering
    \includegraphics[width=0.3\linewidth]{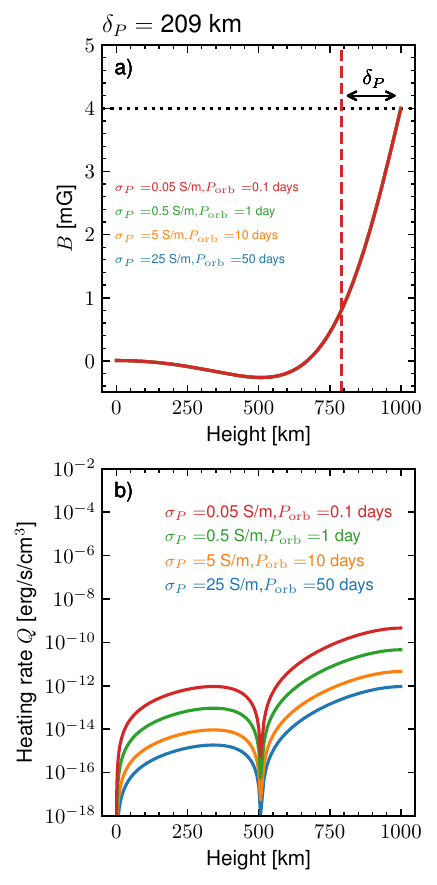}
    \includegraphics[width=0.3\linewidth]{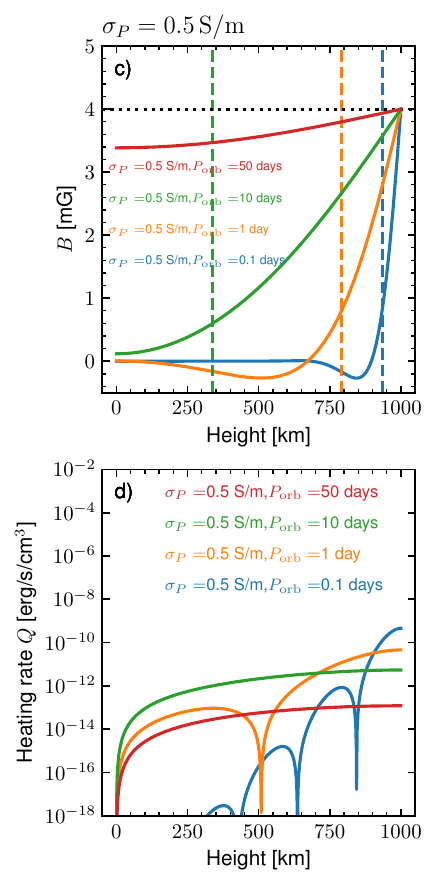}
    \includegraphics[width=0.3\linewidth]{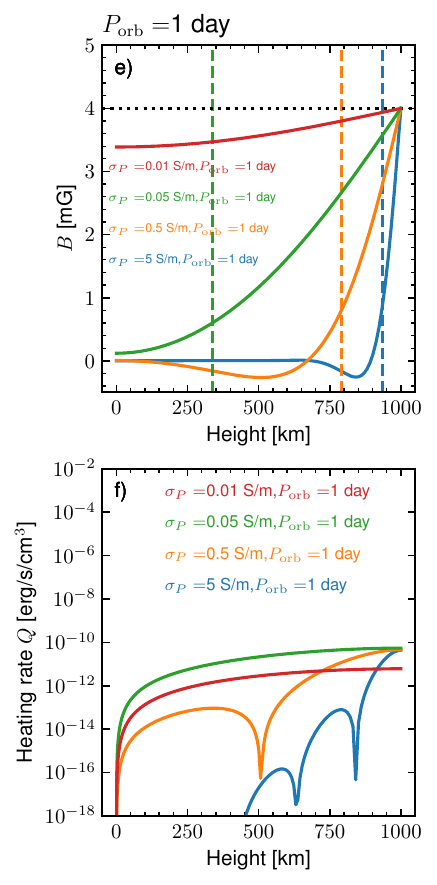}
    \caption{Penetration of time-varying magnetic field $B_{\rm sw} $(panels a, c, e) and associated Ohmic heating $Q$ (panels b, d, f) in the case of a constant $\sigma_P$ atmospheric layer of 1,000 km, subject to a time-varying magnetic field of 4 mG imposed as a boundary condition at the top of the layer. The Hall conductivity is neglected here ($\sigma_H=0$) and the Pedersen conductivity is constant in space. The dotted black line in panels a), c) and e) correspond to the strength of the external time-varying magnetic field $B_{\rm sw}$. Panels a) and b) show cases with constant skin-depth $\delta_P$ and different ($\sigma_p$, $P_{\rm orb}$) as indicated in the legend (note that in panel a, all lines are superposed). In panel a, the skin-depth $\delta_P$ (which is the typical length-scale over which $B_{\rm sw}$ penetrates in the atmosphere) is highlighted. Note that the value of $\delta_P$ in each case is shown by the vertical colored dashed lines in the upper panels (see text). Panel c) and d) depict cases with the same $\sigma_P=$0.5 S/m but different $P_{\rm orb}$. Panels e) and f) show cases with the same $P_{\rm orb}=$1 day but different values of $\sigma_P$.}
    \label{fig:ConstantSigP}
\end{figure*}

\section{Ohmic heating and magnetic screening for simple conductivity profiles}
\label{sec:IllustrativeExamples}

Before moving to more realistic atmospheric profiles, we first illustrate the processes of Ohmic heating and magnetic screening for simplified conductivity profiles.

\subsection{Constant conductivity}
\label{sec:constantSigp}

 We start with the cases of an atmosphere with constant Pedersen conductivity to clearly separate the effects of magnetic screening and Ohmic heating when atmospheres are subject to an external time-varying magnetic field. 

The external time-varying magnetic field can penetrate in the atmosphere over the skin depth $\delta_P$ (Eq. \ref{eq:SkinDepth}), which depends on the oscillation frequency $\Omega=2\pi P_{\rm orb}^{-1}$ of the time-varying field and on the Pedersen conductivity in the atmosphere $\sigma_P$ ($P_{\rm orb}$ is the orbital period of the exoplanet here). As a result, the profile of the oscillating field and of the associated current density depends on this parameter $\delta_P$. The associated heating $Q$ (Eq. \ref{eq:heating}) then depends on this current density as well as the Pedersen and Hall conductivities. It is therefore possible to have atmospheres with the same skin-depth $\delta_P$ but for which the oscillating magnetic field leads to different heating rates. We illustrate this situation in the panels a) and b) of Fig. \ref{fig:ConstantSigP}, for the case of a layer 1,000 km thick, subject to an oscillating field of amplitude $B_{\rm sw}=4$ mG. In panel a) we show the profile of the penetrating field $B$ as a function of height. The four cases depicted in these panels have the same $\delta_P=209$ km (shown by the vertical dashed line), as a consequence the profile of $B_{\rm sw}$ in the atmosphere is the same. Nevertheless, we see in panel b) that the associated heating rate is different. The case with the larger conductivity and larger orbital period (in blue) harbors the weaker heating. The heating of the case with the smaller conductivity (here 0.05 S/m) is stronger by three orders of magnitude. This can be easily understood, since it corresponds to the dissipation of a similar magnetic energy over a shorter timescale due to the smaller $P_{\rm orb}$. 

It is also instructive to highlight the effect of changing only the oscillating period of the external field (panels c and d, constant $\sigma_P=0.5$ S/m) or changing only the Pedersen conductivity  (panels e and f, constant $P_{\rm orb}=1$ day). We have constructed these illustrative examples such that each color corresponds to the same skin-depth in panels c) and e) ($\delta_P$ is shown by the vertical dashed line for the green, orange and blue cases). The blue case corresponds to small skin-depth, while the red case corresponds to a case where the skin-depth is larger than the size of the atmospheric layer considered in this example. 

In panel d) (constant $\sigma_P$ case), we see the maximum heating rate decreases with increasing orbital period. The shape of the heating rate changes as well, because for small skin-depths the field oscillates only over several $\delta_P$ within the atmosphere as it is dissipated efficiently. In the green and red cases, the skin-depth is sufficiently large so that this phenomenon does not take place. 

The case where we keep the oscillation period $P_{\rm orb}$ constant and change instead the Pedersen conductivity is notably different. Indeed, we see that in this situation the maximum heating rate reaches the same value at the top of the atmosphere in the three models having a skin-depth smaller than the size of the atmospheric layer (cases in blue, orange and green). This is remarkable, because for these three cases $\sigma_P$ is varied by two orders of magnitude. The penetration of $B_{\rm sw}$ in the atmosphere leads to a current density ${\bf J}$ that depends on $\sigma_P$. When neglecting $\sigma_H$, the heating rate $Q={\bf J}^2/\sigma_P$ becomes then independent of $\sigma_P$ as long as the skin-depth is smaller than the size of the atmospheric layer (as shown in Appendix \ref{sec:constantPedersenCondCase}). This fact is actually quite physically intuitive: when the oscillation period of the external field is set (which is the case of the problem under consideration), the magnetic energy available to be dissipated is always the same. Changing $\sigma_P$ then changes both the ability of the atmosphere to screen the external field and its ability to dissipate it locally, resulting in a maximum heating rate independent of $\sigma_P$. We note, nonetheless, that because $\delta_P$ changes significantly for these three examples, the profile of heating rate changes significantly as well in the considered atmospheric layer.

Finally, in the illustrative cases presented so far, the maximum heating rate is always found to occur at the top of the layer we consider. This is of course due to the fact that we consider an oscillating field from the top of the atmosphere, and a constant profile of Pedersen conductivity. We now turn to illustrating cases with spatially variable $\sigma_P$, where the maximum heating will not necessarily occur at the top of the atmosphere.

\subsection{Conductivity variations with height}
\label{sec:varSigp}

\begin{figure*}
    \centering
    \includegraphics[width=\linewidth]{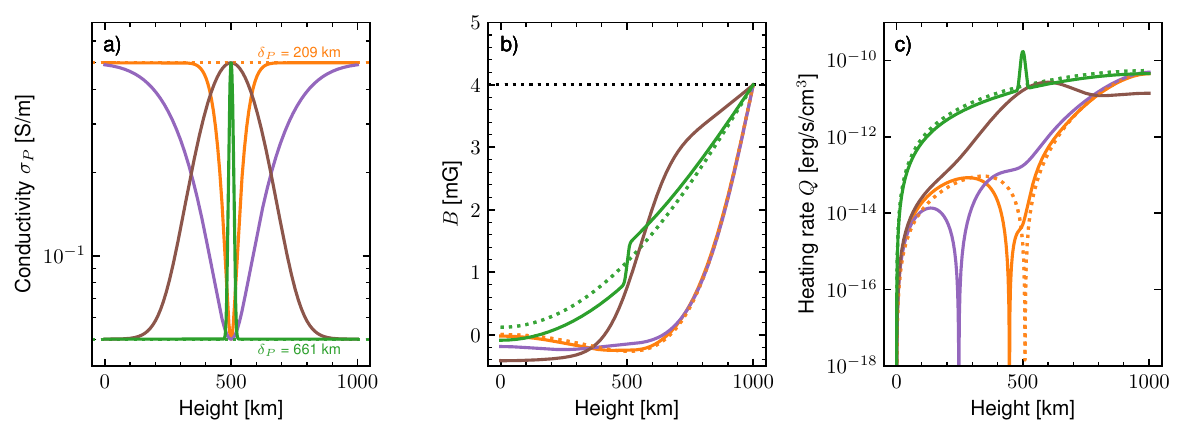}
    \caption{Same experimental setup than in panels e) and f) of Fig. \ref{fig:ConstantSigP} but with a Pedersen conductivity $\sigma_P$ that varies in space. The Pedersen conductivity profiles are shown in panel a). The magnetic field profiles are shown in panel b), and the heating rates in panel c). The dashed orange and green curve correspond to the solid orange and green curves in panels e) and f) of Fig. \ref{fig:ConstantSigP}.}
    \label{fig:VariableSigP}
\end{figure*}

We now illustrate the effect of conductivities that are spatially varying, keeping all other parameters constant. As in Sect. \ref{sec:constantSigp}, we consider a layer of 1,000 km, an external field $B_{\rm sw}=4$ mG, a vanishing Hall conductivity and an oscillating period $P_{\rm orb}=1$ day.

We illustrate four cases where the conductivity is symmetric with respect to the middle of the layer, and either is maximum at the sides or in the center, as shown in panel a) of Fig. \ref{fig:VariableSigP}. The orange and green cases have top and bottom conductivities equal to the conductivity considered in the constant-$\sigma_P$ orange and green cases, shown in the panels e) and f) of Fig. \ref{fig:ConstantSigP}. In the orange case, the conductivity is dropped to a small value in the middle of the layer over a length-scale of about 140 km. We see in panel b) that this drop does not affect much the penetration of the magnetic field, as the profile of $B$ with the drop (solid orange line) and without the drop of conductivity (dashed orange line) are very similar. This is due to the fact that the conductivity at the top of the layer dominates the penetration at the upper end of the atmosphere. Likewise, the heating rate (panel c) is mildly affected by the drop of conductivity. Such a drop is generally considered as providing insulation of the lower atmosphere and planetary interior from the upper atmosphere and interplanetary medium \citep{knierim_shallowness_2022}. We see with this example that it can be very dependent on how strong the drop actually is, and on which length-scale it occurs.

The opposite case is illustrated by the green model: in this case the conductivity is small except in a thin layer of about 20 km in the middle of the atmosphere. This situation is typical of what occurs in planetary upper atmospheres, as will be shown for realistic atmospheric profiles in Sect. \ref{sec:T1b_PMc}. In that case, we observe a sharp change in the profile of $B$ at the location of the conductivity enhancement in panel b. This translates also into a peak in the heating rate at the same location, which corresponds to a local increase by a factor 8 compared to the constant conductivity case (compare the solid and dotted green lines in panel c). 

Finally, we have also explored the effect of varying the layer size on which the conductivity drop and enhancements are made. The purple case in Fig. \ref{fig:VariableSigP} shows a case where the conductivity drop impacts the whole layer. In this case, the profile of $B$ is still mostly set by the conductivity at the top of the layer and small differences are only observed in the heating rate in the bottom part of the layer. Conversely, the brown curves show the case where the conductivity is significantly increased over about 300 km in the middle of the atmosphere. In this case, both the profile of $B$ and the heating rate are strongly affected by the change of conductivity when compared to the case in green. The peak in the heating rate is distributed over a larger fraction of the atmospheric layer, and reaches a smaller value than the case where the conductivity enhancement is applied to a smaller portion of the layer (green curve, panel c).

We can conclude a few general trends from this illustration of the effect of simple changes in the conductivity profiles. First, a localized conductivity drop (orange and purple cases in Fig. \ref{fig:VariableSigP}) do not affect much the profile of the penetrating magnetic field nor the peak of the heat deposition. Second, a localized conductivity enhancement (green and brown cases in Fig. \ref{fig:VariableSigP}) have a strong impact on both the penetrating magnetic field, on the peak of the heat deposition, and on the location of this peak. Interestingly, this remains true even if the conductivity enhancement occurs on a layer smaller than $\delta_P$, as it is the case for the green curves for instance. Therefore, we expect that in realistic atmospheric profiles, the level of magnetic penetration and of Ohmic heating will be mostly determined by the conductivity maxima, and the location of the heat deposition will be affected by the shape of the conductivity profile. 

\section{Application to realistic atmospheres: the cases of Trappist-1b and $\pi$ Men c}
\label{sec:T1b_PMc}

\subsection{Comparison of the conductivity profiles of Trappist-1 b, $\pi$ Men c and the Earth}
\label{sec:EarthComparison}

We now turn to more realistic atmospheric profiles to assess whether external time-varying magnetic field can actually penetrate and/or deposit significant heat in the atmosphere of known exoplanets. In this section, we will focus on the cases of Trappist-1 b and $\pi$ Men c that could harbor water-rich atmospheres \citep{garcia_munoz_heavy_2021,agol_refining_2021}. 

\modif{
The photochemistry in the upper atmosphere of these two planets has been modeled by \citet{garcia_munoz_heating_2023}. The model solves simultaneously the mass-momentum-energy conservation equations in a spherical-shell atmosphere (the only spatial coordinate is the vertical direction, and 3D effects are therefore not taken into account). The atmosphere is irradiated from the top by its host star, which is described by a spectral energy distribution that ranges from X-rays to the far-ultraviolet (say, from a few to 2,500 Angstroms). The stellar flux is attenuated by the atoms and molecules in the atmosphere. The deposition of energy drives a chain of photochemical reactions that in turn dictate the molecular-to-atomic-to-ionic transformation of the gas. The same deposition of energy also drives the dynamics of the gas, which accelerates from essentially hydrostatic conditions to velocities of a few km/s at 2-3 planetary radii above the surface of the planet. The current version of the photochemical model comprises about 150 reactions that connect self-consistently the abundances of the electrons, ions and neutrals in the gas. The model does not include yet the production of secondary electrons in a self-consistent way (see \citealt{gillet_self-consistent_2023} for an example of such effect in a hydrogen-dominated atmosphere). 
}

The characteristics of the star and the planet that are important for this work are summarized in Table \ref{tab:PropertiesT1PiMenc}. We will compare our result with the Rosetta stone case of the Earth. The model of the Earth ionosphere is taken from the NRLMSIS-00 model \citep{picone_nrlmsise00_2002} for the neutrals and IRI-2016 for the ions \citep{bilitza_international_2022}. Both models were obtained from the instant request tools available at the Community Coordinated Modeling Center (CCMC) at Goddard Space Flight Center and are computed up to an altitude of 1,000 km. \modif{We note that these models are empirical and therefore differ significantly from the photo-chemical and dynamical model of \citet{garcia_munoz_heating_2023}. They nevertheless constitute an important benchmark for the computation of assessment of conductivities in our study.}

\begin{table}[htbp]
    \centering
    \caption{Characteristics of Trappist-1 and $\pi$ Men systems.}
    \label{tab:PropertiesT1PiMenc}
    \begin{tabular}{c|cc}
         & Trappist-1$^\dagger$ & $\pi$ Men$^\star$ \\
         \hline 
          $M_\star$ [$M_\odot$] & 0.0802 $\pm$ 0.0073 & 1.02 $\pm$ 0.03 \\
          $R_\star$ [$R_\odot$] & 0.117 $\pm$ 0.0036 & 1.10 $\pm$ 0.01 \\
         & & \\
         & Trappist-1 b & $\pi$ Men c \\
         \hline
          $M_p$ [$M_\oplus$] & 0.85 $\pm$ 0.72 &  4.52 $\pm$ 0.81 \\
          $R_p$ [$R_\oplus$] & 1.086 $\pm$ 0.035 &  2.06 $\pm$ 0.03 \\
          $P_{\rm orb}$ [days]   & 1.5108708 ± 6 $\times 10^{-7}$ &  6.26834 $\pm$ 0.00024 \\
          $R_{\rm orb}$ [$R_\star$] & 20.5 & 13.4 \\
            F$_{5-912 \AA}(R_{\rm orb})$  
      & \multirow{2}{*}{9762$^a$} & \multirow{2}{*}{1350$^{b}$}     \\ 
            {[}erg/cm$^{2}$/s{]} 
      &  &         
    \end{tabular}
    \tablefoot{$^\dagger$ Unless stated otherwise, values for the Trappist-1 system have been taken from \citet{gillon_seven_2017}. $^\star$ Unless stated otherwise, values for the $\pi$ Men system have been taken from \citet{gandolfi_tesss_2018}. 
    $^a$ From the MEGA-MUSCLES spectrum of \citet{wilson_mega-muscles_2021}. $^b$ From \citet{garcia_munoz_is_2020}. }
    
\end{table}

\begin{figure*}[!htbp]
    \centering
    \includegraphics[width=0.32\linewidth]{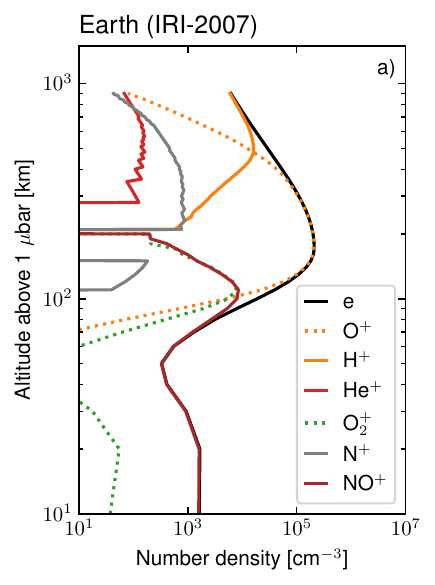}    
    \includegraphics[width=0.32\linewidth]{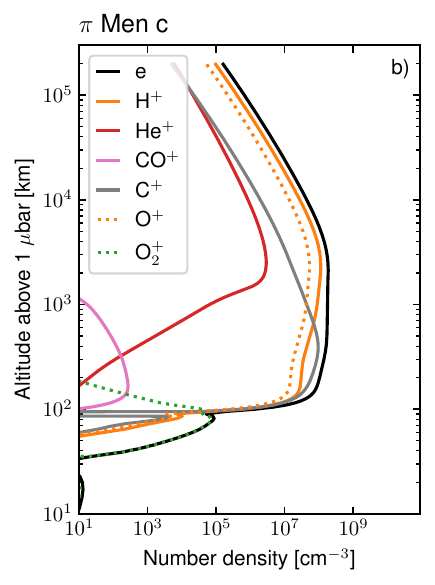} 
    \includegraphics[width=0.32\linewidth]{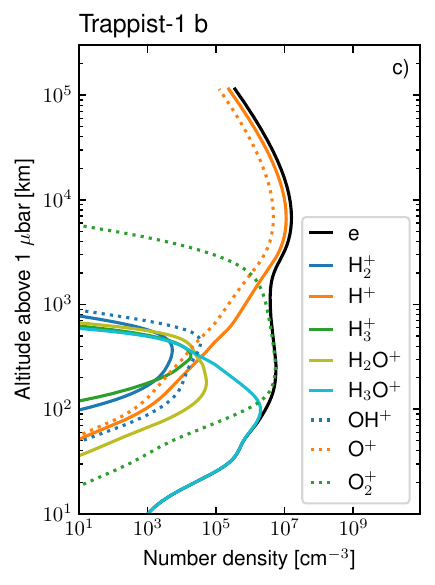}    
    \includegraphics[width=0.49\linewidth]{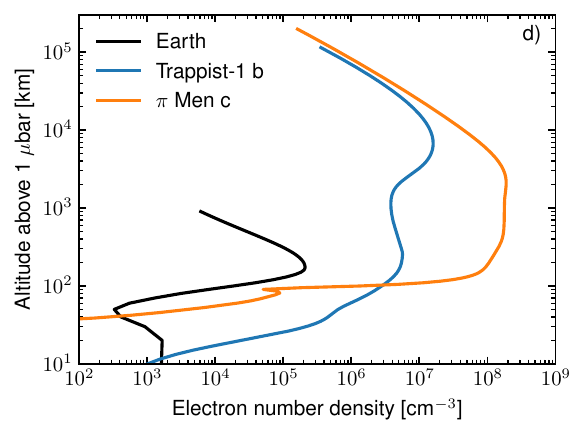}    
    \includegraphics[width=0.49\linewidth]{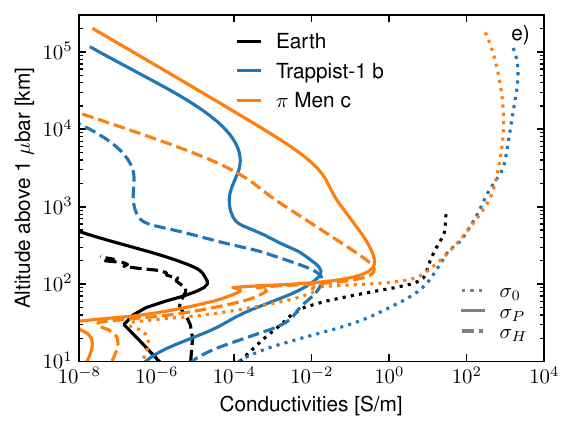}
    \caption{The three upper panels a-b-c present the profiles of the number density of ions (colored lines) and electrons (black line) as a function of height in the upper atmosphere of the Earth (panel a), $\pi$ Men c (panel b) and Trappist-1 b (panel c). Panel d) shows the electron number density of the three models on the same plot. Panel e) shows the conductivity profile above 1$\mu$bar for the Earth (black lines), Trappist-1 b (blue lines) and $\pi$ Men c (orange lines). The parallel conductivity $\sigma_0$ (Eq. \ref{eq:sig0}) is shown by dotted lines. We assume a constant magnetic field $B_P=0.2$ G to compute the Pedersen (solid lines, Eq. \ref{eq:sigP}) and Hall (dashed lines, Eq. \ref{eq:sigH}) conductivities. The underlying atmospheric and ionospheric models are taken from IRI-2016 and NRLMSIS-00 models for the day-side Earth (see text), and \citet{garcia_munoz_heating_2023} for Trappist-1 b and $\pi$ Men c.}
    \label{fig:EarthComparison}
\end{figure*}

The composition of the three atmospheres, the collision frequencies and species contributions to the conductivities (see Sect. \ref{sec:formalism}) are detailed in Appendix \ref{sec:DetailsT1bPMc}. Here we show the electron and ion number densities in the upper panels of Fig. \ref{fig:EarthComparison} for the Earth (panel a), $\pi$ Men c (panel b) and Trappist-1 b (panel c). We first note that the three models have an overall electron number density profile that is similar, but still present some important differences (the electron number densities are also reported in panel d to ease the comparison between the three models). First, due to their proximity to their host star and the strong XUV flux they receive, Trappist-1 b and $\pi$ Men c have electronic densities (black lines in the upper panels) with peak values from one to three orders or magnitude larger than the Earth. The Earth electron number density (black line in panels a and d) presents a peak at an altitude of about 200 km above the 1$\mu$bar level, reaching $n_e \simeq$ 2$\times$10$^{5}$ cm$^{-3}$. At this height, ionic species are dominated by O$^+$ in the Earth atmosphere, and H$^+$ dominates only above (see panel a). $\pi$ Men c (orange line in panel d) has a  different composition (see also Appendix \ref{sec:CompositionT1bPMc}),
with a peak around 200 km linked to C$^+$ ions. These ions are rapidly formed in the deeper layers even though the C/O ratio in the model is $\ll$1 because the ionization potential of the C atom is very low, and far-UV photons generally penetrate deeper than XUV photons. At higher layers, the ionization is dominated by H$^+$ and O$^+$ which produce the observed plateau of $n_e \simeq$ 2$\times$ 10$^{8}$ cm$^{-3}$ (see panel b). Finally, the Trappist-1 b model has the most different atmospheric composition due to a strong abundance of H$_2$O deep in the atmosphere. The electronic number density profile (blue line in panel d) exhibits a double peak structure. The lower peak $n_e \simeq$ 6$\times$ 10$^{6}$ cm$^{-3}$ near 200 km is associated with the presence of H$_3^+$ ions, and the higher peak $n_e \simeq$ 10$^{7}$ cm$^{-3}$ to H$^+$ and O$^+$ ions (see panel c). We stress here that even though the modeled Trappist-1 b receives a stronger XUV flux than the modeled $\pi$ Men c, the latter nevertheless shows more electrons in its upper atmosphere. This is a probably a result of the strong concentration of H$_2$O molecule in the atmosphere of the Trappist-1 model, that leads to a lesser production of electrons despite receiving more flux, as well as a difference in the spectral energy distribution of the stellar spectra. This emphasizes the importance to take into account the composition of the atmosphere to adequately estimate the electron number density, and therefore the conductivity.

The differences in the electron number density lead to significant differences in the parallel, Hall and Pedersen conductivities of the upper atmospheres of the Earth, Trappist-1 b and $\pi$ Men c as shown in panel e of Fig. \ref{fig:EarthComparison}. The same formalism is used to assess the three conductivities for the three extended atmospheres, assuming a constant planetary magnetic field of 0.2 G throughout the atmospheric layer to put ourselves close to Earth-like conditions. We have also compared our computation of the three different conductivities to the Kyoto World Data Center (WDC) for Geomagnetism ionospheric conductivity model\footnote{The WDC ionospheric conductivity model uses the simplified conductivity formulae of \citet{maeda_conductivity_1977} and is available as a web service at \url{https://wdc.kugi.kyoto-u.ac.jp/ionocond/sightcal/index.html}.}, and found a very satisfying agreement with our generic approach (see Appendix \ref{sec:EarthAppendix}).

The conductivity profiles above 1 $\mu$bar are shown in panel e for the day-side of the Earth (black lines), Trappist-1 b (blue lines) and $\pi$ Men c (orange lines). The parallel conductivity is shown by the dotted lines, the Hall conductivity by the dashed lines, and the Pedersen conductivity by the solid lines. We first note that the parallel conductivity (dotted lines) has a similar profile for the three planets. It is low close to 1 $\mu$bar (10$^{-8}$ S/m for $\pi$ Men c, about 10$^{-4}$ S/m for the Earth). Above, it quickly increases to reach 10 S/m at about 100 km in all cases. Higher up, the models of Trappist-1 b and $\pi$ Men c predict that it should keep rising, peaking close to 1,000 S/m. We see that the shape of the parallel conductivity can be much more peaked for strongly irradiated planet than for the Earth. 

The Pedersen conductivity (solid lines) is the most important conductivity to characterize magnetic screening and Ohmic heating, as we have seen in Sect. \ref{sec:problem_formulation}. Again, the overall shape of $\sigma_P$ is roughly similar for the three planets, but its values are very different. The peak of $\sigma_P$ is found to coincide with the inner peak of the electron number density for each atmosphere (see upper panel). For the Earth day side atmosphere it peaks around 10$^{-5}$ S/m. Conversely, $\sigma_P$ peaks at 0.02 S/m for Trappist-1 b and 0.4 S/m for $\pi$ Men c. Considering an external magnetic field varying on a timescale of a day, this leads to a skin depth $\delta_P$ of 30,000 km for the Earth, 1,000 km for Trappist-1 b and 230 km for $\pi$ Men c. Therefore, everything else being equal, it means that the upper atmosphere of the Earth does not screen significantly external magnetic fields varying on a timescale of a day, whereas the upper atmosphere of $\pi$ Men c would screen them almost systematically. In addition, we note that here we considered so far the same planetary field strength $B_P$, the amplitude Pedersen conductivity for a given planet will be modulated by the value of its large-scale magnetic as well. Finally, the Hall conductivities (dashed lines) follow the same trends as the Pedersen conductivities.

\subsection{Magnetic screening and Ohmic heating}
\label{sec:results_T1b_PMc}

The magnetic field of Trappist-1 b and $\pi$ Men c is unknown. Dynamo scaling laws \citep[{e.g.}][]{Christensen2006a} could be generically applied to these objects to estimate their probable magnetic moment. This was done for instance by \citet{McIntyre2019} to a set of detected rocky planets at that time, who predicted a magnetic moment $\mathcal{M}$ of about 0.13 $\mathcal{M}_\oplus$ for Trappist-1 b. Uncertainties are nevertheless large, here we choose to explore a range of realistic planetary field to assess the various possibilities for magnetic screening and Ohmic heating for these two planets. 

\begin{figure*}
    \centering
    \includegraphics[width=\linewidth]{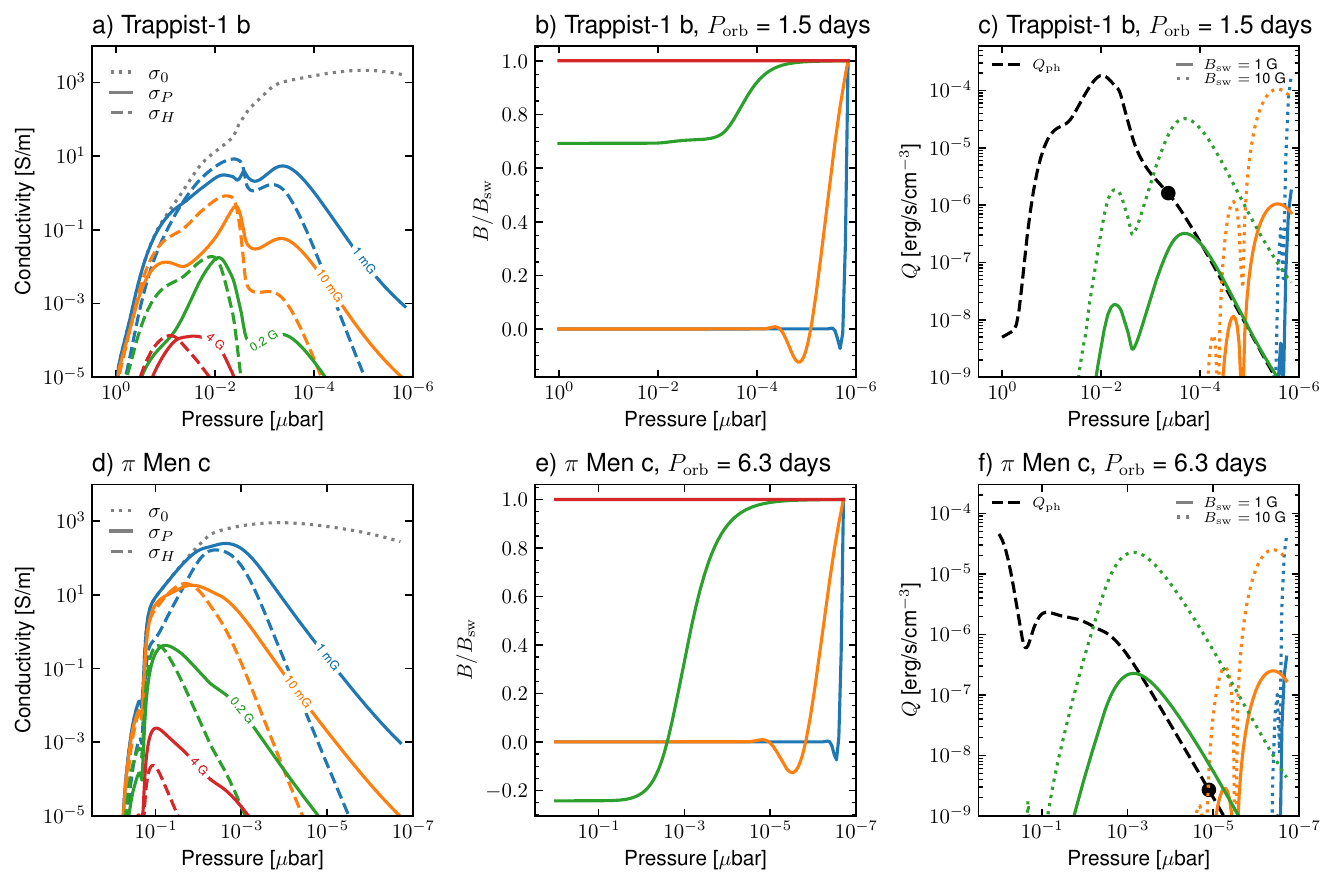}
    \caption{Conductivities, penetration and Ohmic heating in the atmospheres of Trappist-1 b (top panels) and $\pi$ Men c (bottom panels). Left panels (a and d) represent the parallel (dotted gray lines), Hall (dashed colored lines) and Pedersen (solid colored lines) conductivities as a function of pressure in the atmosphere. Four values of planetary magnetic field ($B_P=$ 1 mG, 10 mG, 0.2 G, 4G) are represented in all panels by the blue, orange, green and red lines, respectively. The middle panels (b and e) show the penetrating magnetic field normalized to $B_{\rm sw}$. The right panels (c and f) show the profile of the associated Ohmic heating $Q$ (Eq. \ref{eq:heating}) for $B_{\rm sw}=1$ G  (solid lines) and $B_{\rm sw}=10$ G (dotted lines). The black dashed line show the heating rate from photochemistry in the underlying atmospheric model \citep{garcia_munoz_heating_2023}. Calculations have been done using the orbital period of the planet for Trappist-1 b and $\pi$ Men c (see Table \ref{tab:PropertiesT1PiMenc}).}
    \label{fig:FinalT1bPMc}
\end{figure*}

We first illustrate in panels a) and d) of Fig. \ref{fig:FinalT1bPMc} how the Pedersen and Hall conductivities change assuming a planetary magnetic field of 4 G (Jupiter-like, red lines), 0.2 G (Earth-like, as in fig. \ref{fig:EarthComparison}, green lines), 10 mG (orange lines) and a vanishing magnetic field of 1 mG (blues lines). As expected, the maximum Pedersen conductivity varies by 6 orders of magnitude in the exploration of planetary magnetic fields $B_P$. For each of these profiles, we solve the penetration equation (\ref{eq:FinalMasterEq}) and show the resulting $B_{\rm sw}$ (panels b and e) and Ohmic heating $Q$ (panels c and f, Eq. \ref{eq:heating}) profiles for Trappist-1 b (top panels) and $\pi$ Men c (bottom panels). Here, we have set the oscillation frequency of the time-varying magnetic field to correspond to the orbit of the planet (see Table \ref{tab:PropertiesT1PiMenc}) and assumed an amplitude $B_{\rm sw}=1$G (solid lines) and $B_{\rm sw}=10$G (dotted lines) at the top of the atmosphere.

As we could expect from the results of Sect. \ref{sec:IllustrativeExamples}, a strong planetary field (red curves) leads to a large skin-depth and no effective screening of the external oscillating field. Thus, in this case, both $\pi$ Men c and Trappist-1 b would let the external field penetrate below their atmosphere, without \modif{any interactions between the two}. Conversely, for vanishing magnetic fields (blue curves) the oscillating field is screened at the very top of the atmosphere where its energy is deposited as heat locally. Therefore, even though the conductivity peaks in a relatively narrow region, heating may occur over a much broader range of altitudes in the atmosphere.

In the four illustrated values of $B_P$, the maximum Ohmic heating reaches 2$\times$10$^{-4}$ erg/s/cm$^3$ for Trappist 1-b and 4$\times$10$^{-4}$ erg/s/cm$^3$ for $\pi$ Men c when a very strong $B_{\rm sw}=10$ G field is considered (blue dotted curves in panels c and f). As expected, the Ohmic heating scales with $B_{\rm sw}^2$ and when $B_{\rm sw}=1$ G it decreases by a factor 100 (plain lines). In addition, the amplitude of the planetary field $B_P$ directly dictates the pressure level (or height) in the atmosphere where the Ohmic heating peaks. For instance, in the Trappist-1 b model $Q$ peaks at a pressure level of 10$^{-4}$ $\mu$bar for $B_P=$0.2 G (green lines) and at 2$\times$10$^{-5}$ $\mu$bar for $B_P=$ 10 mG (orange lines). On these same panels c and f, we have added the profile of the photo-chemistry heating rate $Q_{\rm ph}$ modeled by \citet{garcia_munoz_heating_2023} as a black dashed-line. \modif{$Q_{\rm ph}$ does indeed include two separate contributions. These are (i) the heating from the photoelectrons (evaluated as their production rates times their kinetic energies) and (ii) the heating from the fast atoms produced in photo-dissociation processes (similarly evaluated as their production rates times their kinetic energies). The latter contribution is minor at pressures below 0.1 $\mu$bar, but we include it for completeness.}

The predicted Ohmic heating rate generally peaks in the upper part of the modeled upper atmosphere, and can add to and even dominate the photo-chemistry heating rate there. Nevertheless, if the heating rate peaks above the sonic point in the escaping atmosphere, it likely does not affect much the escape process itself nor the overall state of the atmosphere. The sonic point is indicated on the dashed black lines by the black circle. It is relatively close to the peak of $Q_{\rm ph}$ in Trappist-1 b, but is higher up in the atmosphere for $\pi$ Men c. As a result, for strong ambient fields as shown by the dotted colored lines in Fig. \ref{fig:FinalT1bPMc}, the peak of the Ohmic heating is generally further away from the sonic point for Trappist-1 b, but can peak well below the sonic point for $\pi$ men c. In both cases, we see that the green curves can be of comparable amplitude of $Q_{\rm ph}$ below the sonic points, and can in theory therefore participate to the thermal budget and thermal escape of the atmosphere.   

\begin{figure*}
    \centering
    \includegraphics[width=\linewidth]{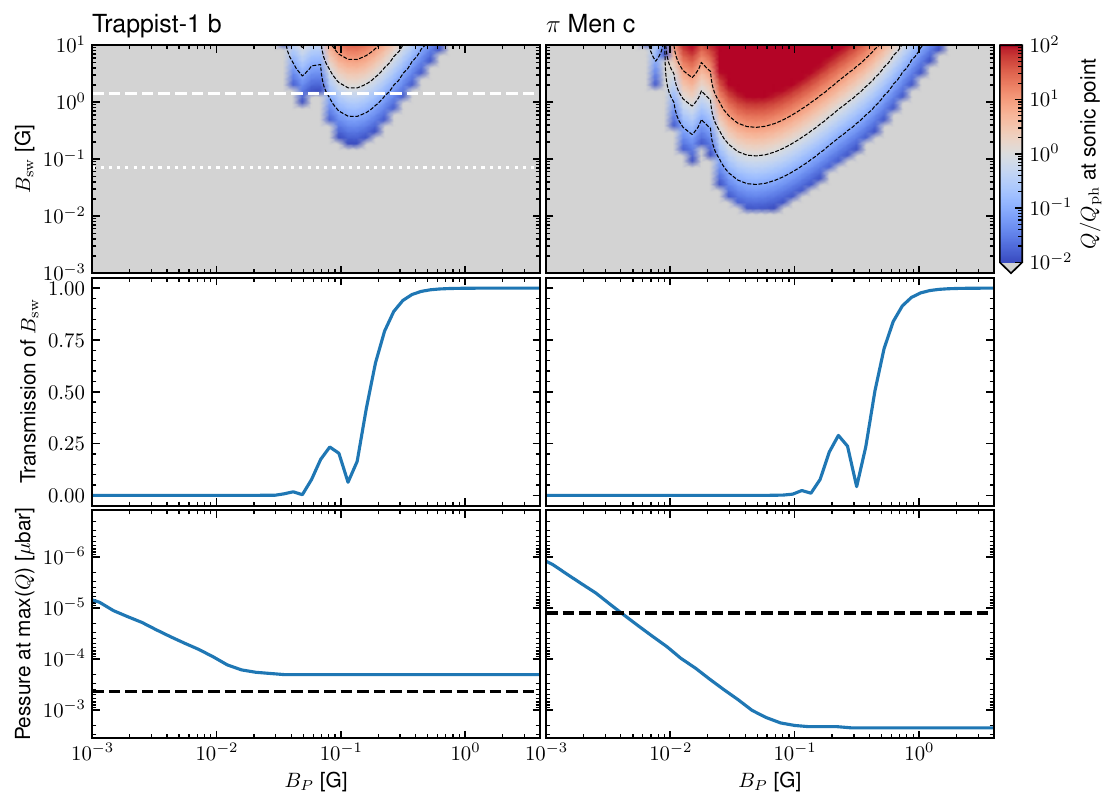}
    \caption{Ohmic heating and field transmission in the upper atmospheres of Trappist-1 b (left column) and $\pi$ Men c (right column). The upper panels show the  heating rate $Q$ at the sonic point $r_c$ divided by the heating from photochemistry $Q_{\rm ph}$ at the same location, as function of the planetary field $B_P$ and the stellar wind field $B_{\rm sw}$. Blue corresponds to $(Q/Q_{\rm ph})(r_c)=0.01$ and red to $(Q/Q_{\rm ph})(r_c)=100$. Values below 0.01 are shown in gray. The dashed and dotted white lines in the top left panel correspond to estimates of the ambient magnetic field based on observations of the magnetic field at the surface of Trappist-1 \citep{reiners_volume-limited_2010}. Note that we have no such constraints for $\pi$ Men at the moment. The thin dashed lines correspond to contours of the ratio of 0.1, 1 and 10. The middle panels show the transmission of $B_{\rm sw}$ from the top of the atmosphere to the level of 1 $\mu$bar as a function $B_P$ (1 corresponds to a full transmission, 0 to a full screening). The bottom panels show the altitude (in pressure value) at which the maximal Ohmic heating rate is realized as a function of $B_P$. The pressure at the sonic point is shown by the black dashed line.}
    \label{fig:TrendsT1bPMc}
\end{figure*}

We can go further and characterize for these two systems the heating rate at the sonic point $r_c$ as a function of $B_P$ and $B_{\rm sw}$ for Trappist-1 b and $\pi$ Men c, as shown by the ratio $(Q/Q_{\rm ph})(r_c)$ in the two upper panels of Figure \ref{fig:TrendsT1bPMc}. We explore the range from 1 mG to 10 G for $B_{\rm sw}$  and 1 mG to 4 G for $B_P$, and display the ratio in a logarithmic scale between 0.01 (blue) and 100 (red). Grey areas corresponds to $Q(r_c) < 0.01 Q_{\rm ph}(r_c)$, {i.e.} regions where the Ohmic heating can be safely ignored. We first note that the qualitative trends are similar in the two panels. The Ohmic heating can be important in the thermal budget for intermediate values of $B_P$, and only if $B_{\rm sw}$ is strong enough. In the middle panels, we show the transmission of $B_{\rm sw}$ (0 means complete screening of the ambient field, 1 perfect transparency) which depends only on $B_P$. We see that if $B_P$ is small enough, the ambient field is completely screened. If $B_P$ is strong enough, it passes through the atmosphere without being dissipated. In the bottom panels, we show the height (in pressure value) at which the heating rate peaks in the atmosphere (blue line), along with the pressure at the sonic point (dashed black line). In the model of Trappist-1 b (left panel), Ohmic heating always peaks above the sonic point whereas in the model of $\pi$ Men c (right panel), it peaks at the sonic point for $B_P\simeq 0.04 G$ and peaks below for larger $B_P$. These lead to the fact that if $B_P$ is small, even if the upper atmosphere screens completely the surrounding field, it does not lead to substantial heating at the sonic point and below (gray areas in the left parts of the upper panels). In that case, Ohmic heating can therefore be safely ignored in the thermal and dynamical budget of upper atmospheres. 

Overall, we can first conclude here that Ohmic heating is expected to be important for the thermal budget of the atmosphere only on a range of intermediate $B_P$ values, when $B_{\rm sw}$ is strong enough. In the case studied here, we predict that for Trappist-1 b it matters if $B_P \in [0.08,0.2] G$, and in the case of $\pi$ Men c if $B_P \in [0.03,0.1] G$, approximately. Note that these values of the planetary field are typically in the range of magnetic moment estimated through dynamo scaling laws for rocky exoplanets by \citet{McIntyre2019}.

In the top left panel of Figure \ref{fig:TrendsT1bPMc}, we have indicated the likely values of $B_{\rm sw}$ for Trappist-1 as dashed and dotted white lines. These values have been deduced from estimates of the surface stellar magnetic flux $ B_\star f$ reported in \citet{reiners_volume-limited_2010} and agree with the 3D wind modeling of Trappist-1 carried out by \citet{reville_magnetized_2024}. We have extrapolated these values assuming a dipolar field $B_{\rm sw} = B_\star f (R_\star/r)^3$ (dotted lines) or a purely radial magnetic flux tube $B_{\rm sw} = B_\star f (R_\star/r)^2$ (dashed lines). This gives $B_{\rm sw}$ between 0.07 and 1.4 G for Trappist-1 b. We have not indicated such an estimate for $\pi$ Men c, as no observational constraints are available on its magnetic field yet to the best of our knowledge. We can therefore conclude that for an ambient field varying on a timescale of the orbital period, Ohmic heating could be important for Trappist-1 b based on the known constraints on the magnetic field of Trappist-1. This is likely true as well as $\pi$ Men c, as weaker magnetic fields compared to Trappist-1 would still lead to a net visible effect of Ohmic heating. This is due to the fact that model of $\pi$ Men c atmosphere presents more electrons than Trappist-1 b, which leads to larger Pedersen conductivities and therefore more efficient Ohmic heating. We note that this conclusion does not necessarily holds for other timescales related to {e.g.} eruptive stellar activity, as will be discussed in Sect. \ref{sec:discussion}.

\section{Application to the exoplanet population}
\label{sec:exoplanetPop}

\subsection{Favorable conditions for magnetic screening and Ohmic heating in the upper atmosphere of hot exoplanets}

Based on the exploration presented above, we can highlight several aspects that can lead to strong heating and magnetic screening in the upper atmosphere of exoplanets. 

In Sect. \ref{sec:T1b_PMc} we have shown that the shape and amplitude of the Pedersen conductivity determines largely the ability of the upper atmosphere to screen and dissipate ohmically an external oscillating field. The Pedersen conductivity is determined by the ions and electrons mixture in the atmosphere (Appendix \ref{sec:DetailsT1bPMc}), as well as the amplitude of the planetary magnetic field in the atmosphere. At first order, the number of electrons in the atmosphere therefore largely determines the level of Pedersen conductivity, as plotted in Fig. \ref{fig:EarthComparison}. Then the detailed ionic composition can also modulate this conductivity. Indeed, we observe that the peak of the Pedersen conductivity differs by a factor around 22 for the atmospheres considered (Fig. \ref{fig:EarthComparison}). The peak of the electron number density differs by a factor around 12 (fig. \ref{fig:Ions}), which explains the order of magnitude of the Pedersen conductivity difference between the two atmospheres. The remaining difference can be understood from the atmospheric composition difference: the Pedersen conductivity of Trappist-1 b is dominated by the contribution from H$_2$O$^+$ and O$_2^+$, whereas in $\pi$ men c it is dominated by H$^+$, C$^+$ and O$^+$ (see Appendix \ref{sec:ConductivitiesT1bPMc}). These levels of electronic and ion density are set by the XUV flux the planet is exposed to and the composition of the atmosphere itself.  

Then, the properties of the time-varying external field determine the accessible energy to be dissipated. The oscillation frequency of the field, coupled to the profile of the Pedersen conductivity, sets the skin-depth ($\delta_P$, Eq. \ref{eq:SkinDepth}) that  determines where in the atmosphere the magnetic field could be ohmically dissipated and at which rate. Then, the amplitude of the externally varying magnetic field determines the amount of magnetic energy that can be dissipated. 

In what follows, we propose to estimate simply the maximal heating rate $Q_{\rm max}$ (see Appendix \ref{sec:constantPedersenCondCase}) which is independent of the conductive properties of the planetary atmosphere. We defer for future work a more in-depth analysis of the population using simplified atmospheric models to assess conductivity profiles and the location of the maximal Ohmic heating.

\subsection{Application to the known exoplanet population}

We estimate the maximal heating rate $Q_{\rm max}$ (Eq. \ref{eq:maxQ_simplified}) that can be awaited due to an oscillation of the ambient magnetic field around the planet at a timescale corresponding to the orbital period. To assess the amplitude of the magnetic field at the planetary orbit, we consider a purely radial magnetic field decreasing away from the stellar surface $B_{\rm sw} = B_\star (R_\star/a)^2$. We estimate the stellar magnetic field based on the 'upper bound' scaling law described in \citet{Ahuir2020} (see their table 1) such that 
\begin{equation}
    B_\star =  \left(\frac{Ro}{Ro_\odot}\right)^{-1} \left(\frac{M_\star}{M_\odot}\right)^{-1.76}\, {\rm [G]}\, ,
\end{equation}
where $Ro=P_{\rm rot}/\tau_c$ is the Rossby number, $\tau_c$ the convective turnover time taken from \citet{lu_abrupt_2023} and See et al. (in prep), and $M_\star$ the stellar mass. In the formula above, $Ro_\odot=0.33$ and $M_\odot$ is the solar mass. A stellar magnetic field of 1 G is obtained for the Sun, which roughly is on par with the solar dipole. Finally, we use as an oscillation frequency $\Omega = 2\pi/P_{\rm orb}$.

\begin{figure*}
    \centering
    \includegraphics[width=\linewidth]{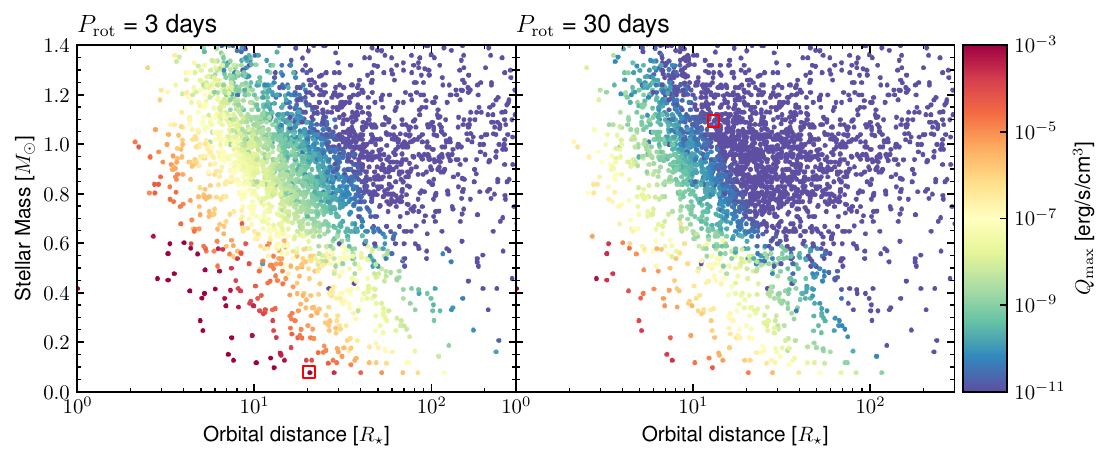}
    \caption{$Q_{\rm max}$ (Eq. \ref{eq:maxQ_simplified}) for the population of exoplanets from the \href{https://exoplanet.eu/}{exoplanet.eu} database, as a function of the orbital distance and the stellar mass. A given rotation rate has been assumed for all stars in each panel, to estimate their magnetic field and the ambient magnetic field strength at the orbit of the planet. Trappist-1 b and $\pi$ men c are highlighted by the red square in the left and right panels, respectively.}
    \label{fig:ExoplanetPop}
\end{figure*}

Based on these parametrization, we can apply our formalism to the whole exoplanet population known as of today. We use the \href{https://exoplanet.eu/}{exoplanet.eu} database. We apply this procedure for two different stellar rotation periods (3 and 30 days). The chosen rotation periods were selected to be close to the rotation period of Trappist-1  \citep[3.3 days,][]{luger_planet-planet_2017}, and a slow rotator that is likely to be representative of a star like $\pi$ Men \citep{gandolfi_tesss_2018}. The resulting $Q_{\rm max}$ is shown for each planet in the database as function of the stellar mass and the orbital distance in the two panels of Fig. \ref{fig:ExoplanetPop}. Trappist-1 b and $\pi$ Men c are highlighted by the red squares in left and right panels, respectively.

This first, crude estimation shows that for close-in planets around low-mass stars, Ohmic heating rates can theoretically reach very high values up to a few 10$^{-3}$ erg/cm$^{3}$/s. The simplified approach presented in this section only allows to estimate the maximal heating rate that can be theoretically achieved. It cannot predict if such level of heating is actually achieved, nor where it occurs in the planetary atmosphere. Those two actually depend on the profile of the Pedersen conductivity in the planet atmosphere, which depends in a non-trivial way on the planetary magnetic field, the composition of the planetary atmosphere, and the XUV and far-UV fluxes it receives. Therefore, a more complete modeling (as done in Sect. \ref{sec:T1b_PMc}) is needed to better assess which close-in planets are likely to have strong Ohmic heating rates affecting their thermal budget below the sonic point of their wind. This will be addressed in a future work.

We finally note that a different formalism was proposed by \citet{cohen_heating_2024}. Their formalism predicts, in general, heating rates that are two to three orders of magnitude larger than the heating rate presented here. This originates from the fact that in their formalism, the maximal heating rate increases \modif{linearly} with the Pedersen conductivity $\sigma_P$. Nevertheless, as $\sigma_P$ increases, the skin depth decreases and rapidly becomes smaller than the size of the layer of interest. In our self-consistent approach, this leads to a saturation of the volumetric heating rate that becomes independent of $\sigma_P$ \citep[see also][]{chyba_internal-current_2021} and equal to $Q_{\rm max}$ (see Eq. \ref{eq:maxQ_simplified}, Appendix \ref{sec:constantPedersenCondCase}\modif{, and Appendix \ref{sec:C24_comparison} for a detailed comparison to the model of \citealt{cohen_heating_2024}}). In any case, this $Q_{\rm max}$ can never be surpassed and can therefore be considered as the correct upper limit for the Ohmic dissipation of external, time-varying magnetic fields.

\section{Discussions}
\label{sec:discussion}

In this work we have considered the screening and dissipation of an external magnetic oscillating on the orbital timescale. Nevertheless, the temporal variations of an external field can also be occurring on shorter and longer timescales. For instance, stellar transients such as coronal mass ejections (CMEs) carry large magnetic fluxes that vary on a crossing-time of the planet. Solar CMEs reach speeds up to 3,000 km/s, which makes them sweep an Earth-like planet in a few seconds to a few minutes. This timescale is typically much shorter than the orbit of the planet, and could therefore lead to even stronger Ohmic dissipation. Because these events are transients and not regular oscillators that we have considered in this work, their effect on the heating of the upper atmosphere must be considered in the more general framework of time-varying field, which will be addressed in future work. Conversely, temporal variations on a larger time-scale can occur due to {e.g.} stellar magnetic cycles. These vary on timescales longer than years \citep[{e.g.}][]{Strugarek2017,Strugarek2018,Brun2022}, making them unlikely to lead to any significant Ohmic heating.

In addition, we have chosen here to neglect the advective component of the induction equation and to consider only time-varying magnetic fields as sources of currents. In the context of a purely radial outflow from the planet, this is likely a valid approximation in the upper dayside of the planetary atmosphere. Deeper in the atmosphere, strong horizontal atmospheric winds are expected to exist. They can induce horizontal magnetic fields from a pre-existing dipolar field, leading to currents as well. The dissipation of these currents, that are not associated with a time-varying magnetic field, also lead to Ohmic heating there \citep[{e.g.}][]{batygin_inflating_2010,rogers_magnetic_2014}. The interplay between these winds and the temperature-dependent conductivity of hot Jupiter atmospheres can even sustain strong temporal variability with episodic bursts of heating \citep{hardy_variability_2022}. We have shown here that even without considering this layer deeper in the atmosphere, the upper atmosphere has the potential to efficiently screen external time-varying magnetic fields. Such fields have been invoked in the past as possible sources of heat deposition in the interior of close-in planets \citep[{e.g.}][]{kislyakova_magma_2017}, and could also add to the net heating in the deep atmosphere. Our study shows that the upper atmosphere screens these external fields in some circumstances, notably when the XUV flux received by the planet is strong enough to lead to large electron densities in the upper planetary atmosphere. Indeed, in that case, the Pedersen conductivity is generally sufficiently high to lead to a small penetration skin-depth. Therefore, our results highlight the need to consider Ohmic heating and induction from the top of the atmosphere down to the planetary interior to properly identify where and how much external time-varying magnetic fields can be ohmically dissipated by planets.


\section{Conclusions}
\label{sec:conclusion}

In this work we have proposed a simplified 1D model for the penetration of external oscillating magnetic fields in the upper atmosphere of exoplanets, and of the associated Ohmic heating. We derived an analytical solution when the conductivity of the atmosphere is constant, and provide an upper limit for the Ohmic heating. Finally, we have also proposed a formalism to apply this approach to multi-species models of the upper atmospheres of exoplanets. We have also shown that in the case of spatially-varying conductivities (which is the typical situation for the upper atmosphere of exoplanets), the conductivity profile can a priori be estimated but the maximum in the conductivity profile does not necessarily matches with the location where Ohmic heating peaks.

We have applied our approach to two iconic hot low-mass planets \citep{garcia_munoz_heating_2023}: Trappist-1 b and $\pi$ Men c. We have showed that due the different XUV flux these planets receive and the differences in their atmospheric composition, $\pi$ Men c likely harbors much stronger upper atmosphere conductivities than Trappist-1 b. Therefore, we find that Trappist-1 b can screen less efficiently external time-varying magnetic fields than $\pi$ Men c. Conversely, both planets can have substantial Ohmic heating that participated to the thermal budget of their upper atmosphere, provided the ambient oscillating magnetic field is strong enough (about 1 G for Trappist-1 b, and about 0.1 G for $\pi$ Men c). 

The modeling presented in this work also allows to characterize where Ohmic heating can occur in the upper atmosphere. This location is primarily set by the existence and strength of the steady magnetic field, which we have assumed to be dominated by the planetary field $B_P$ in this work. If the planetary magnetic field is small, the dominant conductivity is large and the Ohmic heating and magnetic screening occurs high in the atmosphere. For hot exoplanets like Trappist-1 b and $\pi$ Men c, this leads to heating occurring out of the sonic point of the escaping atmosphere. As a result, in that case the planet atmosphere efficiently screens external time-varying magnetic fields, but this has no significant impact on the properties of it escaping atmosphere. Conversely, if the planetary magnetic field is strong the Pedersen conductivity is very small, and the atmosphere is not able to screen and dissipate the external time-varying magnetic field. For intermediate planetary magnetic fields (between $\simeq$0.01 and $\simeq$1 G for the two planets studied in this work), the profile of the conductivity is such that Ohmic heating can occur close to or below the sonic point of the escaping atmosphere. In that case, Ohmic heating should be taken into account self-consistently in the thermal budget of the atmosphere to determine its properties and assess qualitatively by how much the associated atmospheric mass loss rate can change. 

Finally, we have also estimated the maximal Ohmic heating awaited for the population of exoplanets known as of today. We have highlighted that close-in planets around fast-rotating young and low-mass stars are likely susceptible to strong Ohmic heating in their upper atmosphere. Nevertheless, this estimate provides only an upper limit, and dedicated studies on the conductivity profile of these planets is now required. 

The simplified approach presented in this work needs to be extended to be more realistic. First, we have neglected so far inductive effects that should be taken into account. Second, we have considered only perfect oscillators for the external magnetic field. Realistic temporal variations are likely more complicated as they stem from complex magnetic topologies of star, transient events, and eccentric and/or inclined planetary orbits. Third, we have only considered the case of a constant planetary magnetic field throughout the atmospheric layer. The magnetic field most likely diminishes with height, which will make the conductivity increase even more in the upper atmosphere and could amplify the screening and Ohmic heating ever more. We intend to improve these aspects and compare dynamically the effect of such Ohmic heating on models of upper atmospheres of hot exoplanets in a forthcoming work. 

\begin{acknowledgements}

A.S. acknowledges funding from the European Research Council project ExoMagnets (grant agreement no. 101125367) and from the r\'egion Ile-de-France through the DIM Origins project DynamEarths. We acknowledge funding from the Programme National de Planétologie (INSU/PNP). A.S. and A.S.B. acknowledges funding from the European Union’s Horizon-2020 research and innovation programme (grant agreement no. 776403 ExoplANETS-A) and the PLATO/CNES grant at CEA/IRFU/DAp. A.P. acknowledges financial support from the MERAC fundation. 
We acknowledge the Community Coordinated Modeling Center (CCMC) at Goddard Space Flight Center for the use of the Instant Runs tools for the IRI-2016 (\url{https://kauai.ccmc.gsfc.nasa.gov/instantrun/iri/}) and NRLMSIS-00 (\url{https://kauai.ccmc.gsfc.nasa.gov/instantrun/nrlmsis/}) models. This research has made use of data obtained from the portal \href{https://exoplanet.eu/}{exoplanet.eu} of The Extrasolar
Planets Encyclopaedia.

\end{acknowledgements}

%

%


\begin{appendix} 

\section{Composition, collision frequencies and conductivities in the atmospheres of Trappist-1 b and $\pi$ Men c}
\label{sec:DetailsT1bPMc}

\subsection{Composition}
\label{sec:CompositionT1bPMc}

\begin{figure}[!htbp]
    \centering
    \includegraphics[width=\linewidth]{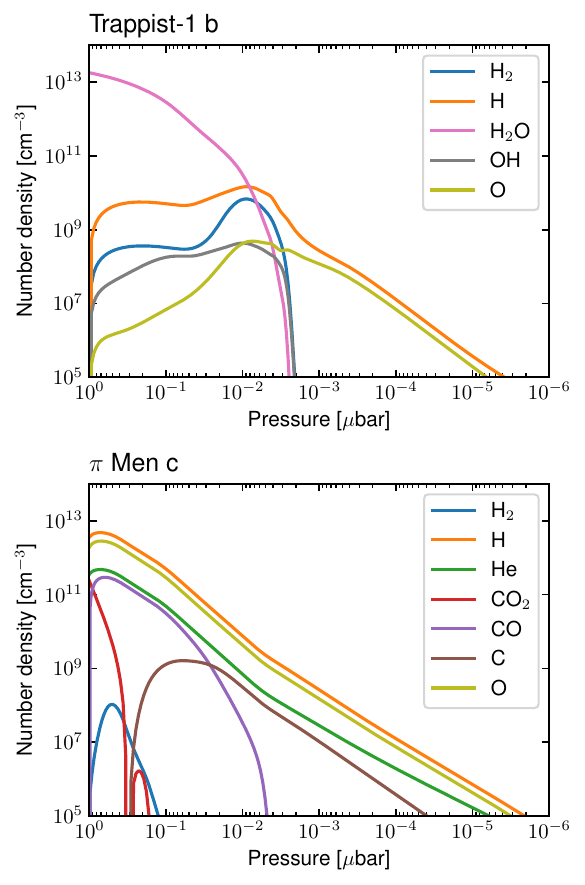}
    \caption{Neutral number densities as a function of height (expressed here as pressure) for Trappist-1 b (top panel) and $\pi$ Men c (bottom panel).}
    \label{fig:Neutrals}
\end{figure}

\begin{figure}
    \centering
    \includegraphics[width=\linewidth]{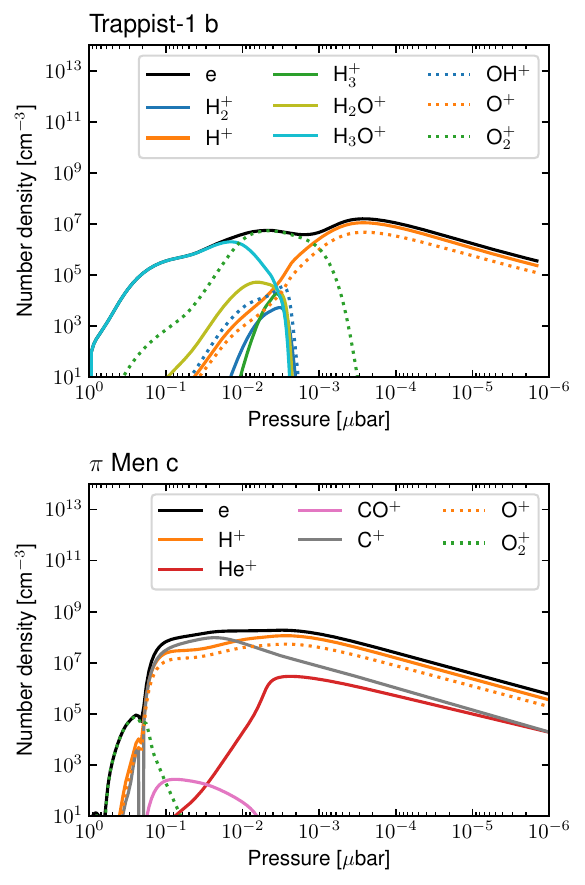}
    \caption{Ion number densities as a function of height (expressed here as pressure) for Trappist-1 b (top panel) and $\pi$ Men c (bottom panel). The electron density is shown by the black line.}
    \label{fig:Ions}
\end{figure}

The photochemistry modeling of the upper atmosphere (pressures above 1 $\mu$bar) of Trappist-1 b and $\pi$ Men c has been conducted by \citet{garcia_munoz_heating_2023}. We use the results of their modeling here to assess the conductive properties of these two upper atmospheres. The number density of neutrals and ions are shown in Fig. \ref{fig:Neutrals} and \ref{fig:Ions}, respectively, as a function of the thermal pressure.

The model of the atmosphere of Trappist-1 b assumed a composition of a 100\% H$_2$O at the pressure level of 1 $\mu$bar. The number density of water vapor decreases from a few 10$^{13}$ cm$^{-3}$ down to about 10$^{10}$ cm$^{-3}$ at 2 nbars where the neutral hydrogen becomes the dominant neutral species in the atmosphere (see top panel of fig. \ref{fig:Neutrals}). Higher in the atmosphere of Trappist-1 b, hydrogen and oxygen are the dominant neutral species. Conversely, in the model of $\pi$ Men c (bottom panel) H$_2$O is unstable at 1 $\mu$bar and therefore its neutral atmosphere is dominated by hydrogen and oxygen at all altitudes.  

The ion composition of the two atmospheres is radically different from 1 $\mu$bar to about 0.5 nbar (see fig. \ref{fig:Ions}). In Trappist-1b, because of the high concentration of water the lower part of the atmosphere is dominated by H$_3$O$^+$ and O$_2^+$ (top panel). At these altitudes, $\pi$ Men c (bottom panel) possesses also a dominant O$_2^+$ deeper down but H$^+$, C$^+$ and O$^+$  quickly dominate the ions higher up. In Trappist-1 b H$^+$ and O$^+$ dominate also in the upper part of the atmosphere, but C$^+$ is completely negligible. 

These two different composition lead to different electron number densities (black lines in fig. \ref{fig:Ions}) when comparing the models of Trappist-1 b and $\pi$ Men c. $\pi$ men c reaches electron densities around 40 times larger than Trappist-1 b. This is somewhat surprising, because $\pi$ Men c receives an XUV flux about 7 times smaller than Trappist-1 b. It shows strikingly that composition, and in particular the fact that the model of Trappist-1 b atmosphere is dominated by H$_2$O determines the level of electron number density. These differences determine the collision frequencies and conductivities that will differ from one atmosphere to the other, which we now turn to.

\subsection{Collision frequencies}
\label{sec:CollisionFreqsT1bPMc}

\begin{figure}
    \centering
    \includegraphics[width=\linewidth]{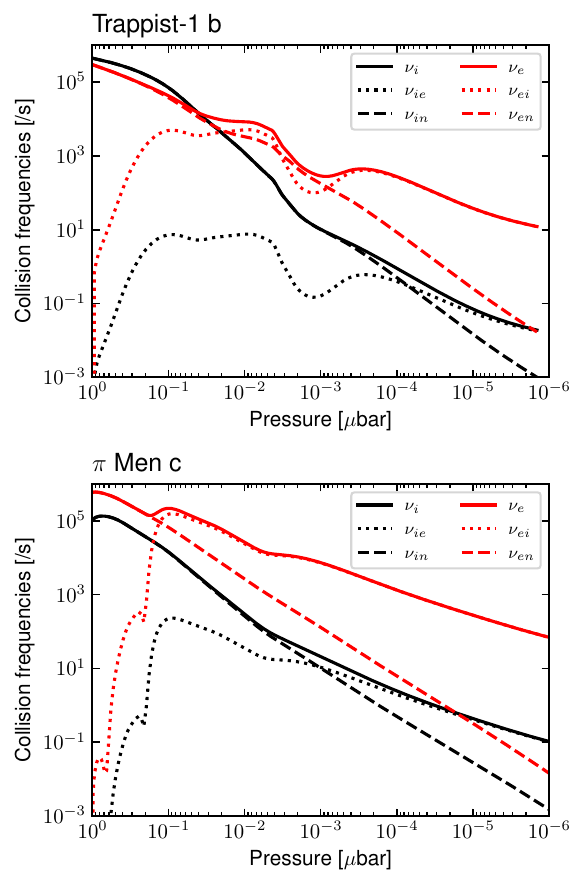}
    \caption{Collision frequencies [/s] for Trappist-1 b (top panel) and $\pi$ Men c (bottom panel) as a function of pressure in the atmosphere.}
    \label{fig:CollFreq}
\end{figure}

We apply the formulae presented in Sect. \ref{sec:CollFreqs} to the case of Trappist-1 b and $\pi$ Men c. The collision frequencies are shown in Fig. \ref{fig:CollFreq}. The total ions (black) and electrons (red) collision frequencies are shown in solid lines. The ions-neutrals and electrons-neutrals collision frequencies are shown by the dotted lines, and the ions-electrons and electrons-ions collision frequencies by the dashed lines.

Collisions are dominated by electrons in the upper atmosphere, and more specifically electrons-ions collisions. For Trappist-1 b, the collisions with neutrals dominates in almost the whole atmosphere for ions. Deeper in the atmosphere, the electron collisions are dominated by the neutrals. 

The collision frequencies are higher at all height for $\pi$ Men c than for Trappist-1 b. We can expect that this leads to larger conductivities in the atmosphere of $\pi$ Men c, which are detailed in the next section. 

\subsection{Conductivities}
\label{sec:ConductivitiesT1bPMc}

\begin{figure*}
    \centering
    \includegraphics[width=\linewidth]{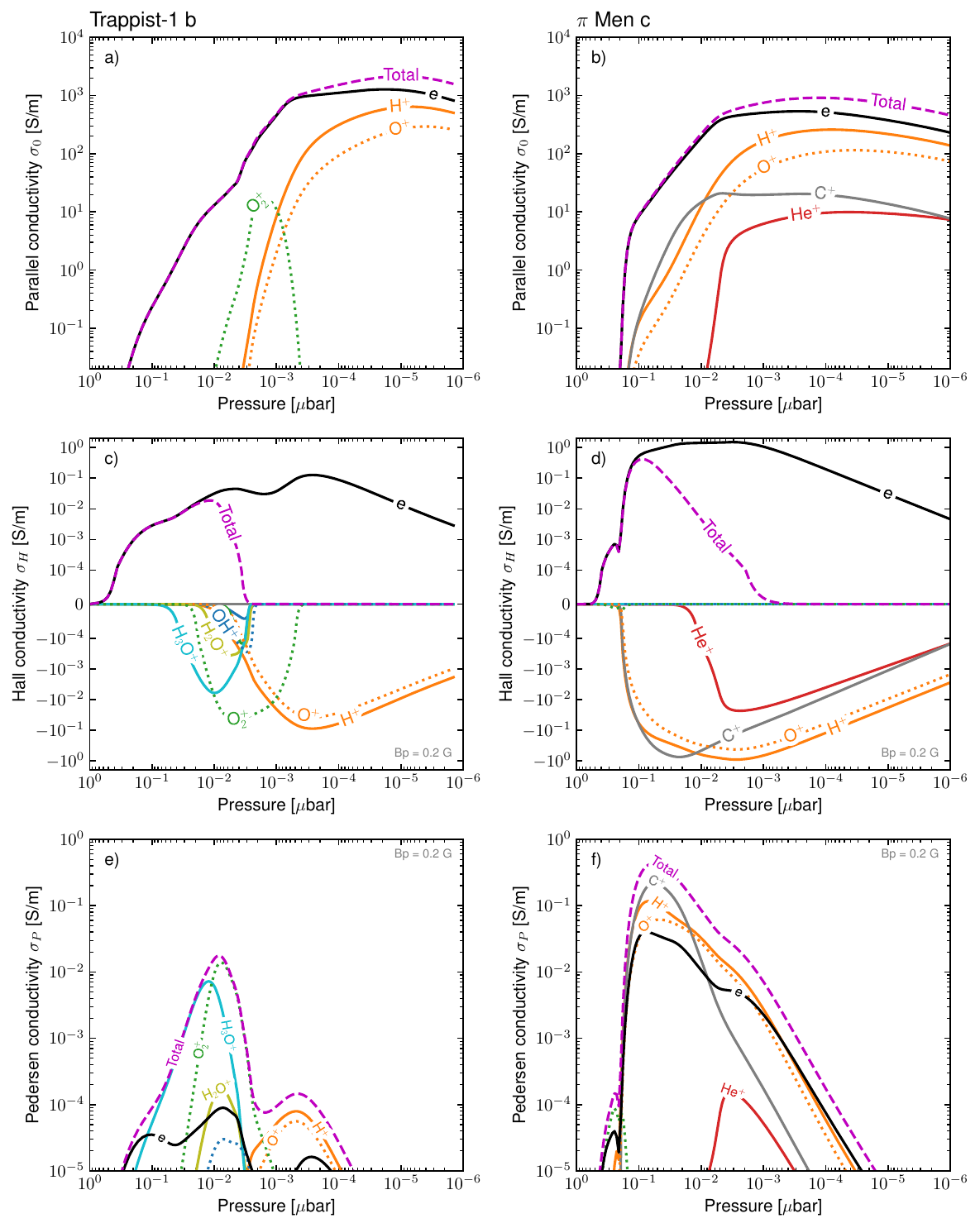}
    \caption{Parallel (top panel), Hall (middle panels) and Pedersen (bottom panels) conductivities for Trappist-1 b (left column) and $\pi$ Men c (right column). The contribution of electrons is shown in black solid lines, the individual contributions of ions in colored solid and dotted lines, and the total conductivity is shown in dashed magenta in each panel. For the Hall and Pedersen conductivity, a constant planetary field of 0.2 G has been assumed throughout the upper atmosphere.}
    \label{fig:SigComp}
\end{figure*}

The parallel, Hall and Pedersen conductivities have been calculated following the formulae laid out in Sect. \ref{sec:ConductivityFormulae}. The resulting conductivities are shown in Fig. \ref{fig:SigComp} as dashed magenta lines. The individual contributions are shown in black (electrons) and colored plain and dotted lines (ions). 

We note first that the parallel conductivity (panels a and b) is mostly determined by the collision of electrons with neutrals in the lower part of the atmosphere and with ions in the upper part of the atmosphere. In both atmospheres, H$^+$ and to a lesser extent $O^{+}$ also contribute directly to the total parallel conductivity. The parallel conductivity is found to be slightly higher in the upper part of the atmosphere of Trappist-1 b than in $\pi$ Men c, and the situation reverses in the lower part of the atmosphere (see also panel d in Fig. \ref{fig:EarthComparison}).

The Hall conductivity peaks around 0.01 $\mu$bar in Trappist-1 b and 0.1 $\mu$bar in $\pi$ Men c. In lower parts of the atmosphere, the electrons dominate again the Hall conductivity. Above the peak, ions compensate the contribution of electrons and render the Hall conductivity negligible. In Trappist-1 b (panel c), O$_2^+$, O$^+$ and H$^+$ are the dominant species affecting the Hall conductivity. In $\pi$ Men c, C$^+$, O$^+$ and H$^+$ are the dominant ones. Overall, the Hall conductivity is higher in $\pi$ Men c than in Trappist-1 b in the upper part of the atmosphere, and smaller in the lower part of the atmosphere.

Finally, as expected, the Pedersen conductivity is dominated by the conductivity associated with ions, and a small contribution from electrons in the lower layers of the considered atmosphere. The Pedersen conductivity peaks at about the same heights as the Hall conductivity in each case. In Trappist-1 b, the Pedersen conductivity is carried from bottom to top by H$_2$O$^+$, O$_2^+$, and then a similar contribution from O$^+$ and H$^+$. In $\pi$ Men c, the Pedersen conductivity is larger due to a significant contribution from C$^+$, H$^+$ and O$^+$. In the upper layers, electrons appear also to provide a significant contribution to the Pedersen conductivity in the atmosphere of $\pi$ Men c.
 
\FloatBarrier



\section{Validation on the case of the Earth}
\label{sec:EarthAppendix}

\begin{figure}
    \centering
    \includegraphics[width=0.85\linewidth]{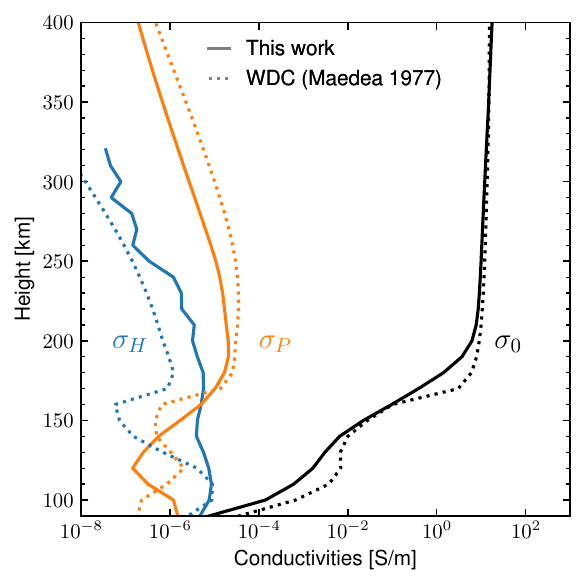}
    \caption{Comparison of the conductivity profiles of the Earth. The profiles obtained from the WDC ionospheric conductivity model (\url{https://wdc.kugi.kyoto-u.ac.jp/ionocond/sightcal/index.html}) are shown as dotted lines. The profiles obtained with the formulation derived in Sect. \ref{sec:problem_formulation} are shown as solid lines. The field-free conductivity $\sigma_0$ (Eq. \ref{eq:sig0}) is shown in black, the Pedersen conductivity (Eq. \ref{eq:sigP}) in orange, and the Hall conductivity (Eq. \ref{eq:sigH}) in blue.}
    \label{fig:ConductivityValidationEarth}
\end{figure}

To validate the approach presented in this work, we compare the calculation of the parallel, Pedersen and Hall conductivities with the profiles obtained with the formulae of \citet{maeda_conductivity_1977} and available in the WDC ionospheric conductivity model at  \url{https://wdc.kugi.kyoto-u.ac.jp/ionocond/sightcal/index.html}. To do so, we have downloaded for the same dates and resolution the profiles for neutrals (NRLMSIS-00, \citealt{picone_nrlmsise00_2002}), ions (IRI-2016, \citealt{bilitza_international_2022}), and the conductivity profiles from the WDC cited above. We apply the formalism presented in Sect. \ref{sec:ConductivityFormulae} to calculate the conductivities, here for a constant planetary field of 0.2 G. The results are shown in Fig. \ref{fig:ConductivityValidationEarth}. The solid lines present the calculations carried with our approach, and the dotted lines are the calculations obtained with the online tool at WDC. The model for the Earth magnetic field is not specified on the website, which makes it hard to exactly reproduce their results. We note, though, that the parallel conductivity (black lines) agree remarkably well, with small differences stemming from the different formulations chosen for the collision frequencies in each approach. The Pedersen conductivities (orange lines) are also similar, to a satisfactory level in the context of the present study that aims at characterising the order of magnitude of Ohmic heating. The Hall conductivity (blue lines) are also similar, yet they present some discrepancies especially near 150 km. These differences likely originate from a combination of different formulations for the collision frequencies, simplified formulae of Pedersen and Hall conductivities considered by \citet{maeda_conductivity_1977}, as well as the fact that the WDC model likely considers a variation of the amplitude and direction of the planetary magnetic field $B_P$ with height, which we have ignored for now. Overall, the conductivities obtained with our approach and with the WDC ionospheric conductivity model are  similar, which validates our current approach for the work presented here.

\FloatBarrier

\section{Analytical Ohmic heating in the case of a constant Pedersen conductivity}
\label{sec:constantPedersenCondCase}

In the case of a constant Pedersen conductivity $\sigma_P$ (and therefore a constant skin-depth $\delta_P$), the problem of magnetic field penetration and associated Ohmic heating can be solved analytically. We derive this solution here and provide a fully analytical formula for this limit. 

Following \citet{Parkinson}, we introduce the generic decomposition
\begin{equation}
    A_0 = A_k \exp^{-ikx} \, .
\end{equation}
Applying this decomposition to Eq. \ref{eq:FinalMasterEq}, we find that $k$ admits two values which are
\begin{eqnarray}
    k = \pm \frac{1}{\delta_P}\left(1+i\right)\, .
\end{eqnarray}
As a result, solutions to Eq. \ref{eq:FinalMasterEq} take the form
\begin{equation}
    A_0 = A_1 e^{-\frac{x}{\delta_P}\left(1-i\right)} + A_2 e^{\frac{x}{\delta_P}\left(1-i\right)}\, .
\end{equation}
Considering the boundary conditions that at the top of the domain ($x=x_t$), the field matches the stellar wind field ($\partial_x A_0 = B_{\rm sw}$) and at the bottom of the domain ($x=x_b$) there is no net current ($\partial_{xx}A_0=0$), we find the unique solution to this problem to be
\begin{eqnarray}
    \label{eq:UniqueSolutionAnalytical}
    A_1 &=& - \frac{B_{\rm sw} \delta_P}{2} e^{\frac{2x_b-x_t}{\delta_P}\left(1-i\right)}\left(1+i\right)\left[ 1 + e^{2\frac{(x_b-x_t)}{\delta_P}\left(1-i\right)} \right]^{-1} \, , \\
    A_2 &=&  \frac{B_{\rm sw} \delta_P}{2} e^{-\frac{x_t}{\delta_P}\left(1-i\right)}\left(1+i\right)\left[ 1 + e^{2\frac{(x_b-x_t)}{\delta_P}\left(1-i\right)} \right]^{-1} \, .
    \label{eq:UniqueSolutionAnalytical2}
\end{eqnarray}

These formulae give the unique solution to the problem at stake. Nonetheless, it is instructive to assess the net heating associated with this solution in the case where the skin depth $\delta_P$ is smaller than the extent of the layer $d=x_t-x_b$. In that case, the denominator of $A_1$ and $A_2$ is dominated by the exponential and the imaginary part of $A_0$ can be estimated to be
\begin{equation}
    \Im(A_0) \simeq \frac{B_{\rm sw} \delta_P}{2}\, .
\end{equation}
This leads to an estimate of $Q$ following Eq. (\ref{eq:heating}) to be 
\begin{equation}
    \label{eq:simplifiedHeating}
    Q^{\delta_P\ll d} \simeq \frac{\sigma_p}{\sigma_H^2+\sigma_P^2} \left(\frac{c}{4\pi}\right)^2 \frac{ B_{\rm sw}^2 }{\delta_P^2}\, .
\end{equation}

In this limit, we remark that because $\delta_P\propto \sigma_P^{-1/2}$, if we neglect the Hall conductivity the Ohmic heating rate becomes essentially independent of $\sigma_P$ to be
\begin{equation}
    \label{eq:final_simplified_ohmicHeating}
    Q_{\rm max} =\frac{\Omega B_{\rm sw}^2}{8\pi} \, .
\end{equation}
This simple formula (in cgs units) can also be obtain assuming that the magnetic energy density contained in the external magnetic field is dissipated over a timescale $\Omega^{-1}$. A similar finding was obtained in spherical geometry by \citet{chyba_internal-current_2021} (see their Eq. 6). This sets the maximal volumetric heating rate that can be achieved by the dissipation of an external oscillating magnetic field.

\modif{In the limits where the skin-depth $\delta_P$ is larger than the size of the layer $d$, the solution is strongly affected by the choice of boundary conditions at the bottom of the layer (here $\partial_{xx}A_0=0$). Within this limit, performing a Taylor expansion of (\ref{eq:UniqueSolutionAnalytical}-\ref{eq:UniqueSolutionAnalytical2}) leads to the fact that $A_0$ becomes proportional to $\sigma_P$. As a result, the estimate of $Q$ using Eq. (\ref{eq:heating}) leads to 
\begin{equation}
    Q^{\delta_P\gg d} \simeq \left(\frac{\sigma_P}{\sigma_P^0}\right)^3 Q_{\rm max}\, ,
    \label{eq:simplified_large_deltaP}
\end{equation}
where we have introduced the Pedersen conductivity that corresponds to $\delta_P=d$, $\sigma_P^0 = c^2/(d^2 2\pi \Omega)$. This limit differs from the limit obtained when considering the induction in a full sphere \citep[{e.g.}][]{Parkinson,chyba_magnetic_2021}, where the internal boundary condition notably differs from the one used in this work. Nevertheless, in this limit the external oscillating magnetic field goes through the layer without much interaction, and therefore does not play much role in the thermal budget of the upper atmosphere.
}

\section{Comparison to the model of Cohen et al. (2024)}
\label{sec:C24_comparison}

\begin{figure}
    \centering
    \includegraphics[width=\linewidth]{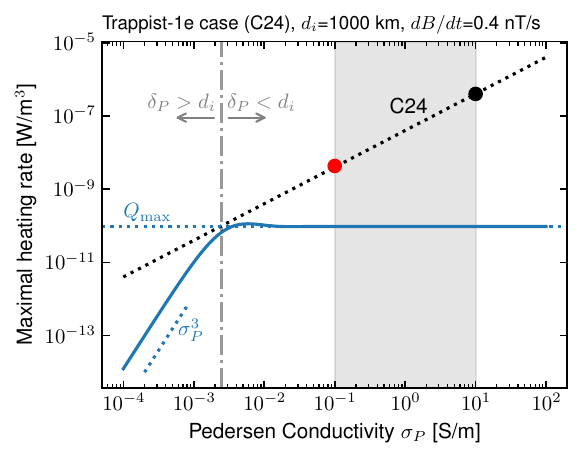}
    \caption{\modif{Comparison to the modeling of \citetalias{cohen_heating_2024} for their case of Trappist-1 e, i.e. for a layer $d_i=$1000 km and a magnetic field variation $dB/dt=$ 0.4 nT/s. We made use of their Eq. 13 to trace $Q$ as a function of $\sigma_P$ (black dotted line). The red and black circles correspond to the red and black circles in their Figure 4 (in the case $d_i$=1000 km). The gray area labels the range of application considered in \citetalias{cohen_heating_2024}. The results of our model are shown by the blue line. The maximal heating rate $Q_{\rm max}$ is labeled by the dotted blue line. The vertical dash-dotted gray line labels the Pedersen conductivity $\sigma_P^0$ for which $\delta_P=d_i$.}
    }
    \label{fig:CompareC24}
\end{figure}

\modif{
In this appendix we compare the results of the formalism used in this paper with the results of \citet{cohen_heating_2024} (hereafter \citetalias{cohen_heating_2024}). To do so, we selected one case from \citetalias{cohen_heating_2024}: the case of Trappist-1 e where the authors consider a magnetic field variation $dB/dt=$0.4 nT/s and an ionospheric layer of thickness $d_i=$ 1000 km. We use their Equation 13 to estimate the volumetric heating rate using their formalism. The resulting heating rate is shown as a function of $\sigma_P$ in Fig. \ref{fig:CompareC24} as a dotted black line (labeled C24). We have reported the values found in their figure 4 for $\sigma_P=0.1$ S/m (red circle) and $\sigma_P=10$ S/m (black circle) divided by the layer thickness $d_i$. 
}

\modif{
In order to apply their parameters with our formalism, we used the analytical solution from Appendix \ref{sec:constantPedersenCondCase} using 
\begin{equation}
    \Omega^{C24} = \frac{dB/dt}{B} \, ,
\end{equation}
with $dB/dt=0.4$ nT/s and $B=600$ nT from \citetalias{cohen_heating_2024}. We have also tried using $B=2000$ nT (maximum magnetic field along the orbit of Trappist-1 e in \citetalias{cohen_heating_2024} model), which does not change the results presented here. The results of our models are shown as a blue solid line in Fig. \ref{fig:CompareC24}.
}

\modif{We find first that when $\sigma_P=\sigma_P^0$ (dash-dot vertical gray line), both models agree and predict a volumetric heating rate close to $Q_{\rm max}$. For smaller $\sigma_P$ values, the models predict a decreasing $Q$ with different slopes, owing to the choice of boundary conditions (see end of Appendix \ref{sec:constantPedersenCondCase}). For $\sigma_P>\sigma_P^0$, the two models predict strikingly different values for the volumetric heating rate. The model of \citetalias{cohen_heating_2024} linearly scales with $\sigma_P$, whereas our self-consistent approach predicts that $Q$ should saturate at $Q_{\rm max}$ (as in {e.g.} \citealt{chyba_magnetic_2021}), leading to drastically smaller values of $Q$ in the range of conductivities considered by \citetalias{cohen_heating_2024} (gray area). We interpret this discrepancy by the fact that in the model of  \citetalias{cohen_heating_2024}, the skin-depth effect is not taken into account. This effect is nevertheless very important, as the external oscillating field should be screened over the skin-depth when it is able to trigger a current that is dissipated by the atmosphere.
}

\end{appendix}
\end{document}